\documentclass[fleqn,usenatbib]{mnras}

\usepackage{newtxtext,newtxmath}
\usepackage{pifont}
\usepackage{anyfontsize}

\usepackage[T1]{fontenc}

\DeclareRobustCommand{\VAN}[3]{#2}
\let\VANthebibliography\thebibliography
\def\thebibliography{\DeclareRobustCommand{\VAN}[3]{##3}\VANthebibliography}

\usepackage{sty/support-caption}
\usepackage{subcaption} 
\usepackage{graphicx}	
\usepackage{amsmath}	
\usepackage{xcolor}     

\usepackage{comment}
\def\mistral    {{\sc mistral}}
\def\mistralC    {{\sc mistral-continuous}}
\def\mistralS    {{\sc mistral-stochastic}}
\def\nobh       {{\sc no bh}}
\def\tng       {{\sc fiducial tng}}
\def\quasar       {{\sc tng thermal}}
\def\isoth       {{\sc isotropic thermal}}
\def\randomw     {{\sc random kinetic}}
\def\arepo    {{\sc arepo}}
\def\subfind    {{\sc subfind}}
\def\sublink    {{\sc sublink}}
\def\ngenic    {{\sc N-GenIC}}
\def\gadget    {{\sc gadget}}
\def\gizmo    {{\sc gizmo}}

\def\smuggle    {{\sc smuggle}}
\def\fire    {{\sc fire}}
\def\ramses    {{\sc ramses}}

\newcommand\MF{\textcolor{red}}

\title[The Mistral AGN feedback model]{MISTRAL: a model for AGN winds from radiatively efficient accretion in cosmological simulations}

\author[Farcy et al.]{
Marion Farcy,$^{1}$\thanks{E-mail: marion.farcy@epfl.ch}
Michaela Hirschmann,$^{1}$
Rachel S. Somerville,$^{2}$
Ena Choi,$^{3}$
Sophie Koudmani,$^{2,4,5,6}$
\newauthor
Thorsten Naab,$^{7}$
Rainer Weinberger,$^{8}$
Jake S. Bennett,$^{9}$
Aklant K. Bhowmick,$^{10}$
Hyunseop Choi,$^{11,12}$
\newauthor
Lars Hernquist,$^{9}$
Julie Hlavacek-Larrondo,$^{11}$
Bryan A. Terrazas,$^{13}$
Francesco Valentino$^{14,15}$
\\
$^{1}$Institute of Physics, Laboratory for Galaxy Evolution, EPFL, Observatoire de Sauverny, Chemin Pegasi 51, 1290 Versoix, Switzerland\\
$^{2}$Center for Computational Astrophysics, Flatiron Institute, 162 5th Avenue, New York, NY 10010, USA\\
$^{3}$Department of Physics, University of Seoul, 163 Seoulsiripdaero, Dongdaemun-gu, Seoul 02504, Republic of Korea\\
$^{4}$St Catharine’s College, University of Cambridge, Trumpington Street, Cambridge CB2 1RL, UK\\
$^{5}$Institute of Astronomy and Kavli Institute for Cosmology, University of Cambridge, Madingley Road, Cambridge, CB3 0HA, UK\\
$^{6}$Centre for Astrophysics Research, Department of Physics, Astronomy and Mathematics, University of Hertfordshire, College Lane, Hatfield, AL10 9AB, UK\\
$^{7}$Max-Planck-Institut für Astrophysik, Karl-Schwarzschild-Strasse 1, 85741 Garching, Germany\\
$^{8}$Leibniz Institute for Astrophysics Potsdam (AIP), An der Sternwarte 16, 14482 Potsdam, Germany\\
$^{9}$Center for Astrophysics $\vert$ Harvard \& Smithsonian, 60 Garden St, Cambridge, MA 02138, USA\\
$^{10}$Department of Astronomy, University of Virginia, Charlottesville, VA 22904, USA\\
$^{11}$Département de Physique, Université de Montréal, Succ. Centre-Ville, Montréal, Québec, H3C 3J7, Canada\\
$^{12}$Mila - Quebec Artificial Intelligence Institute, Montréal, Quebec, Canada\\
$^{13}$Department of Physics \& Astronomy, Oberlin College, Oberlin, OH, 44074, USA\\
$^{14}$Cosmic Dawn Center (DAWN), Denmark\\
$^{15}$DTU Space, Technical University of Denmark, Elektrovej 327, DK-2800 Kgs. Lyngby, Denmark}

\date{Accepted XXX. Received YYY; in original form ZZZ}

\pubyear{2025}

\begin{document}
\label{firstpage}
\pagerange{\pageref{firstpage}--\pageref{lastpage}}
\maketitle

\begin{abstract}
Feedback from active galactic nuclei (AGN) is crucial for regulating galaxy evolution. Motivated by observations of broad absorption line winds from rapidly accreting supermassive black holes (SMBHs), we introduce the \mistral{} AGN feedback model, implemented in the \arepo{} code. \mistral{} comes in two versions: continuous radial (\mistralC{}) and stochastic bipolar momentum deposition (\mistralS{}). Using the framework of the IllustrisTNG simulations, we explore the effect of \mistral{} on BH and galaxy properties, through an idealized Milky Way-mass galaxy and cosmological zoom simulations run down to $z=2$. Unlike standard thermal AGN feedback prescriptions, \mistral{} generates galaxy-scale winds that mimic outflows driven by BH accretion. \mistralC{} produces short-lived galactic fountains, and is inefficient at regulating the growth of massive galaxies at $z=2$. In contrast, \mistralS{} efficiently suppresses star formation in massive galaxies, reproduces the empirical stellar-to-halo mass relation, and yields a consistent trend of BH-stellar mass evolution. By supporting large-scale outflows while simultaneously preventing gas inflows, \mistralS{} additionally regulates the cold and hot gas fractions at both galaxy and halo scales. \mistralS{} therefore works self-consistently across the halo mass range explored $\left(10^{12}-3\times10^{13}\,\rm M_\odot\right)$, without adopting a SMBH-mass dependent AGN feedback scheme such as the one used in IllustrisTNG. Our model is a promising tool for predicting the impact of AGN winds on galaxy evolution, and interpreting the growing population of high-redshift galaxies and quasars observed by JWST. This work is part of the "Learning the Universe" collaboration, which aims to infer the physical processes governing the evolution of the Universe.
\end{abstract}




\section{Introduction}

For more than two decades, most (if not all) massive galaxies are thought to host a supermassive black hole (SMBH) in their centres, with masses ranging from a million to a few billion solar masses \citep{Magorrian1998,Schodel2002}. As material accretes onto SMBHs, vast amounts of energy are released in the form of energetic radiation, winds and jets \citep{Fabian2012,Tombesi2013,Cicone2014,Genzel2014}, powering so-called active galactic nuclei (AGN). Since the energy released by a SMBH can easily exceed the binding energy of its host galaxy \citep[e.g.][]{King&Pounds2015}, AGN are thought to play a significant role in galaxy evolution \citep{Silk&Rees1998,Kauffmann&Haehnelt2000,Bower2006,Croton2006,Somerville2008,Somerville&Dave2015,Naab&Ostriker2017}. Depending on the efficiency with which the BH accretion energy released couples to the surrounding gas, the back reaction of an AGN -- referred to as AGN feedback -- can affect both BH growth and the evolution of the host galaxy by heating and ejecting gas, and (or) preventing gaseous inflows, and thus depriving both star formation and BH accretion of their source of fuel \citep{Fabian2012,Gaspari2012,Harrison2017}. This BH-galaxy coevolution scenario is further supported by the tight correlations observed between the BH mass and the properties of their host galaxies, such as stellar bulge luminosity, mass and stellar velocity dispersion (\citealp{Tremaine2002,Kormendy&Oh2013,Reines&Volonteri2015}, even if also non-causal origins of these relations have been discussed by e.g. \citealp{Hirschmann2010,Jahnke2011}).

In the past decade, observations have found increasingly strong and abundant evidence for AGN feedback, in form of jets and winds. Collimated, radio-emitting jets are thought to be powered by the extraction of the SMBH spin, through the Blandford-Znajek process \citep{Blandford&Znajek1977}. Jets are associated with radiatively inefficient AGN events, when SMBHs are accreting at low rates \citep[e.g.][]{McNamara2000,Fabian2011,Hlavacek-Larrondo2022}. By inflating hot X-ray cavities in the intra-cluster medium of galaxies, jets are commonly invoked as an efficient mechanism for reducing cooling flows onto galaxies, and thus, subsequent star formation, which is also referred to as the radio or jet feedback mode \citep[e.g.][]{Fabian2012}. However, in the radiatively inefficient regime, other forms of feedback can exist, such as outflows, without the presence of strong relativistic jets. Therefore, throughout this paper, we refer to AGN feedback in this low-accretion state as radiatively inefficient feedback.

Conversely, in high-luminosity AGN (with bolometric luminosities of $10^{46}-10^{48}\,\rm erg\,s^{-1}$), the accreting mass is efficiently converted into radiation, driving winds from radiatively efficient accretion disks. These winds are powered through magnetic or radiation pressure on lines, free electrons or dust. They can originate on small scales within the SMBH torus or accretion disk, in the broad line region, or at larger nuclear scales. AGN winds are observed via broad absorption lines (BAL) in the rest-frame UV, as well as broad optical, UV and infrared emission lines in the spectra of quasars \citep[e.g.][]{King&Pounds2015}, which are characteristic of strong outflows of molecular and ionised gas \citep[e.g.][]{Richards2011,Harrison2014,Cicone2014,Hamann2018,Wylezalek2020,Herrera-Camus2020}. In particular, BAL winds, found in $\sim20-50$ percent of the quasar population across all redshifts \citep[e.g][]{Hewett&Foltz2003,Dai2008,Bischetti2023, Maiolino2024}, are considered a key driver of AGN feedback; they are often highly energetic, with outflow velocities ranging from a few thousand to a few tens of thousands of $\rm km\,s^{-1}$ \citep[e.g.][]{Fiore2017,Choi2020,RodriguezHidalgo2020}, and can lead to outflows that extend from parsec to galactic scales, where the latter can reach several kpc from the central SMBH or even further \citep[e.g.][]{Arav2018,Choi2022}.

By transferring mass, momentum and energy from the nuclear region to the interstellar medium (ISM) and the circumgalactic medium (CGM) of galaxies, winds from radiatively efficient accretion expel and heat large amounts of gas, and therefore have the ability to impact star formation \citep[e.g.][for a review]{Veilleux2020}. This AGN feedback mode is thought to be particularly relevant at cosmic noon ($z\simeq2$), when both cosmic star formation density and quasar activity peak \citep{Madau&Dickinson2014}. This scenario is further supported by the prevalence of AGN-driven circum-nuclear outflows observed at this epoch \citep[e.g.][]{Schreiber2014,Genzel2014}, indicating the importance of radiatively efficient AGN feedback (also referred to as quasar-mode feedback in the literature) in the early assembly of galaxies.

The close connection between AGN feedback and galaxy evolution has been explored extensively in semi-analytic models and numerical simulations \citep[for reviews see][]{Somerville&Dave2015,Naab&Ostriker2017}. AGN feedback is a key ingredient of current galaxy formation models, which is necessary to reproduce observable properties of galaxy populations in the nearby to intermediate redshift Universe. This includes the formation of a population of red and dead massive galaxies, the regulation of the massive end of the galaxy stellar mass function, realistically low stellar content at a given halo mass, and the regulation of the cooling rate and X-ray temperature of gas at galaxy cluster scales \citep{Somerville2008,Ciotti2010,Gaspari2011,Dubois2013,Harrison2017,Henden2018,Eckert2021}. However, the details of how these processes actually work remain uncertain, due to the impossibility for simulations to capture the dynamic range from BH to galaxy scales. For instance, fully resolving BH accretion would require reaching physical scales on the order of the Schwarzschild radius. For a BH mass of $10^7\,\rm M_\odot$, this corresponds to $10^{-6}\,\rm pc$, which is computationally prohibitive in cosmological simulations that aim at modelling hundreds of Mpc.

To smooth over the small-scale complexity, the common approach to modelling AGN feedback in cosmological simulations is to rely on subgrid prescriptions, meant to reproduce the consequences of processes that happen at unresolved scales. Initially, AGN feedback was implemented by simply assuming that a fraction of the rest-mass SMBH accreted energy couples thermally to the surrounding gas \citep{Springel2005,DiMatteo2005}, which was shown to be insufficient in reducing cooling and star formation in most massive systems with stellar masses beyond $10^{11}\rm\,M_\odot$ (at the typical resolution achieved in cosmological simulations). To overcome this, more complex prescriptions have been progressively developed \citep{Choi2017,Bourne&Sijacki2017,Talbot2021,Husko2022,Rennehan2024,Husko2024,Husko2025}, though some of these are still prohibitively expensive to include in large cosmological simulations. In such coarse resolution simulations, isotropic thermal BH energy deposition remains a widely adopted approach to implementing feedback in the radiatively efficient BH regime (e.g. \citealp{DiMatteo2012,Vogelsberger2014c,Hirschmann2014,Dubois2014,Springel2018}, but see \citealp{Schaye2015,Choi2015,Dave2019}).

To better regulate gas fractions and star formation in the most massive galaxies, some simulations transition to another AGN feedback mode when the Eddington ratio drops below a certain threshold ($\sim 10^{-2}$), injecting kinetic energy \citep{Dubois2012,Weinberger2017}, or thermal energy with a higher efficiency \citep{DiMatteo2012,Steinborn2015} or in the form of AGN-inflated bubbles of hot gas \citep{Sijacki2007}. Other simulations, which do not distinguish between radiatively efficient and inefficient regimes \citep{Schaye2015,Schaye2023}, distribute the thermal AGN energy from a reservoir of feedback energy by stochastically heating gas elements to a high temperature \citep{Booth&Schaye2009}. To reproduce galaxy observables at $z=0$, cosmological simulations often enhance the effectiveness of their AGN thermal mode, for instance by stochastically heating a selected number of cells to a large enough temperature to overcome excessive radiative losses \citep{Schaye2015}, or implement a two-mode feedback scheme with an accretion rate-based switch \citep{Dubois2012,Weinberger2017,Dave2019}. The latter approach is adopted for instance in the Illustris-TNG simulations \citep[TNG hereafter,][]{Springel2018,Pillepich2018,Naiman2018,Nelson2018,Marinacci2018,Nelson2019}.

Despite adopting a range of different AGN feedback models, cosmological simulations have been successful in reproducing basic observations of galaxy, BH and AGN populations, predominantly for the low-redshift Universe. Such simulations include MassiveBlack \citep{DiMatteo2012}, Illustris \citep{Vogelsberger2014c}, Magneticum \citep{Hirschmann2014}, Horizon-AGN \citep{Dubois2014}, MassiveBlack-II \citep{Khandai2015}, Eagle \citep{Schaye2015}, Romulus \citep{Tremmel2017}, Illustris-TNG \citep{Springel2018}, Simba \citep{Dave2019} and Flamingo \citep{Schaye2023}. However, their galaxy formation models are usually calibrated to reproduce selected galaxy and BH properties at $z=0$, which limits their predictive power for the interpretation of galaxy evolution through cosmic time. In particular, \citet{Habouzit2021,Habouzit2022} showed that predictions for high-redshift BH and AGN populations from cosmological simulations tend to strongly diverge with increasing redshift. Moreover, most current cosmological simulations underestimate the observed number density of high-redshift massive quiescent galaxies \citep[e.g.][]{Valentino2023,Weller2025}.

In addition, it remains unclear whether radiatively efficient AGN feedback in the form of thermal energy deposition can generate realistic galactic-scale gaseous outflows, as observed out to $z=7.5$ \citep[][and references therein]{Liu2024}. Thus, motivated by observations of BAL winds, \citet{Ostriker2010} and \citet{Choi2012,Choi2015} first implemented a momentum/kinetic energy based AGN feedback model for generating winds driven by radiatively efficient BH accretion. Using similarly motivated prescriptions \citep{AnglesAlcazar2017,Dave2019,Costa2020}, recent simulations suggest that quasar-driven outflows can play an important role in galaxies. Notably, they contribute to efficiently regulating star formation and X-ray luminosities \citep{Choi2018,Costa2018,Dave2019}, bringing the $z=0$ galaxy stellar mass function and cold and hot gas fractions in simulations closer to observations, compared to incorporating only continuous thermal quasar energy injection \citep{Springel2005,Sijacki2007}.

Motivated by this success, it is now within reach to investigate how AGN winds from radiatively efficient accretion affect galaxy and BH evolution out to early cosmic epochs. In this paper, we build on the AGN feedback model of \citet{Ostriker2010} and \citet{Choi2012} and introduce the \mistral{}\footnote{Mistral is the name of a cold and strong wind that blows in South of France.} subgrid model for cosmological simulations, implemented into the \arepo{} code \citep{Springel2010}. \mistral{} bridges the gap between radiatively efficient AGN wind prescriptions in high-resolution, idealised simulations \citep[e.g.][]{Costa2020,Sivasankaran2025}, and traditional subgrid models implemented in large volume cosmological simulations. As such, \mistral{} allows us to investigate the role of radiatively efficient AGN-driven winds in galaxy evolution. In particular, in this paper, we use \mistral{} to explore to what extent AGN-driven winds affect BH and their host galaxy properties around cosmic noon.

For this purpose, in this study, we couple \mistral{} with the well-tested galaxy formation physics from the TNG cosmological simulations. For reproducing realistic galaxy populations at $z=0$, the TNG simulations distinguish between a thermal isotropic AGN energy input at high Eddington ratios (referred to as the TNG thermal model), and a kinetic prescription in the low-accretion regime (introduced by \citealp{Weinberger2017}). Our aim with \mistral{} is to incorporate a physically motivated prescription for AGN winds from radiatively efficient accretion, in order to understand their impact on the BH-galaxy co-evolution compared to previous, standard AGN feedback models. This is also one of the key goals of the "Learning the Universe" collaboration\footnote{\url{https://learning-the-universe.org}}, of which this work is a part. With this intention in mind, we have explored two variants of our model implementation: a continuous radial deposition of momentum (\mistralC{}), and stochastic bipolar momentum injection (\mistralS{}) -- the \arepo{} analogue of the model implemented by \citet{Choi2012} in the \gadget{} code \citep{Springel2005gadget}. In this study, both versions are used independently, and at all Eddington ratios.

The paper is organised as followed. Section~\ref{section:models} gives an overview of the galaxy formation and BH physics used in the TNG simulations and in this study (Section~\ref{subsec:physics-tng}), before presenting the details of the \mistral{} model (Section~\ref{subsec:bh-mistral}). Throughout the paper, we study how AGN winds generated by our two versions of \mistral{} impact BH growth and galaxy evolution, compared to the AGN feedback models used in the standard TNG simulations. We split this analysis into two sections. We first explore the effect of \mistral{} within an idealised simulation of a Milky Way-mass galaxy (Section~\ref{section:resultsm12}). To evaluate the ability of \mistral{} to realistically regulate BHs and galaxies, we analyse galaxy and outflow properties at $z=2$ in a suite of 15 cosmological zoom simulations (Section~\ref{section:resultsz2}), whose initial conditions are extracted from the TNG100 simulation.
We finally comment on our results in the context of other studies in Section~\ref{section:disc}, and summarise our conclusions in Section~\ref{section:ccl}.

\section{Galaxy formation and Black Hole physics models}
\label{section:models}

In this section, we first focus on the models for galaxy and BH physical processes used in the TNG project, which we adopt in the simulations in this paper. Then, we present our new \mistral{} feedback model and its implementation into the astrophysical code \arepo{} \citep{Springel2010}.

\subsection{Galaxy formation physics in TNG}
\label{subsec:physics-tng}

Throughout this paper (aside from our new AGN feedback model), we use the same galaxy formation model as in the published TNG simulations, which reproduce several key galaxy properties and scaling relations at $z=0$ \citep[see e.g.][]{Pillepich2018b}. These models are implemented in the magnetohydrodynamical moving-mesh code \arepo{} \citep{Springel2010,Weinberger2020}, and explained in detail in \citet{Weinberger2017,Pillepich2018}. The interstellar medium is modelled with a two-phase, effective equation of state, based on the model developed by \citet{Springel&Hernquist2003}. Radiative gas cooling is computed from pre-tabulated cooling functions, depending on gas density, temperature and metallicity. Stars form stochastichally from gas denser than $n_{\rm H}\simeq0.13\,\rm cm^{-3}$, following the Kennicutt-Schmidt relation and the \citet{Chabrier2003} initial mass function. As they evolve, stars chemically enrich the surrounding gas and release mass and energy as  type Ia and type II supernova and Asymptotic Giant Branch (AGB) stars. Stellar feedback is modelled as supernova-driven isotropic winds, releasing 10\% of the energy thermally, following \citet{Marinacci2014}, and the rest in a kinetic, isotropic fashion. Unlike the model used in the original Illustris simulations \citep{Vogelsberger2014c}, supernova-driven winds in TNG have an energy that depends on the star-forming gas cell metallicity, such that the total wind energy decreases with metallicity (Eq. 3 from \citealp{Pillepich2018}). The stellar-driven outflows are mediated by wind particles, launched at a velocity that scales with the dark matter velocity dispersion. These wind particles are decoupled from the dense, local ISM, and propagate until they encounter a gas cell with a low density (0.05 times lower than the density threshold for star formation), into which they transfer their mass, momentum, metal and energy.

In the TNG simulations, SMBHs are modelled as collisionless sink particles. They are seeded with an initial mass $M_{\rm BH} = 1.2\times10^6\,\rm M_\odot$ in halos with a dark matter mass above $7.4\times10^{10}\,\rm M_\odot$ that have not been previously seeded with a BH. Throughout the simulation, the SMBHs are (re)located at the position of the halo potential minimum every timestep. They can grow by merging with other BHs and by accreting gas, as we explain in more detail in the next section. AGN feedback operates in a dual mode: a thermal mode at high Eddington ratios (the radiatively efficient BH regime, Sec.~\ref{subsubsec:quasar-tng}), and a kinetic mode at low Eddington ratios (radiatively inefficient regime, Sec.~\ref{subsubsec:jet-tng}).


\subsubsection{Black hole accretion}
\label{subsubsec:accretion}

Black hole accretion in the TNG simulations follows the Eddington-limited Bondi-Hoyle-Lyttleton prescription \citep{Hoyle&Lyttleton1939,Bondi&Hoyle1944,Bondi1952}, where the Bondi-Hoyle-Lyttleton accretion rate $\dot{M}_\mathrm{Bondi}$ and the Eddington rate $\dot{M}_\mathrm{Edd}$ are defined as follows:

\begin{align}
    \label{eq:bondi}
    &\dot{M}_\mathrm{Bondi} = \frac{4 \pi G^2 M_\mathrm{BH}^2 \rho}{c_{\rm s}^3}\\
    \label{eq:edd}
    &\dot{M}_\mathrm{Edd} = \frac{4 \pi G M_\mathrm{BH} m_{\rm p}}{\epsilon_{\rm r} \sigma_{\rm T} c}
\end{align}

Here, $G$ is the universal gravitational constant, $m_{\rm p}$ is the proton mass, $\epsilon_{\rm r}$ is the radiative accretion efficiency, $\sigma_{\rm T}$ is the Thompson cross-section, and $c$ is the vacuum speed of light. $M_\mathrm{BH}$ is the black hole mass, and $\rho$ and $c_{\rm s}$ are the average density and effective sound speed\footnote{The effective gas sound speed includes the contribution from the thermal gas speed $c_{\rm s, th}$ and the speed of the Alfven waves $c_{\rm A}$, such that $c_{\rm s}^2=c_{\rm s,th}^2+c_{\rm A}^2$.} of the gas near the black hole, respectively. All the gas properties are calculated within the BH smoothing length $h_{\rm BH}$, by performing a kernel-weighted average over a prescribed number of gas cells $n_{\rm BH,ngb}$ that are the nearest BH neighbours, such that:

\begin{align}
    \label{eq:ngb}
    & n_{\rm BH,ngb} = \sum_i \frac{4\pi h_{\rm BH}^3}{3}\frac{m_i}{m_{\rm target}}w_{\rm k}(r_i)\, ,
\end{align}
where $m_i/m_{\rm target}$ is the mass of the gas cell $i$ compared to the target gas mass resolution enforced by the refinement scheme of the \arepo{} code. $w_{\rm k}(r_i)$ is the cubic spline smoothing kernel weight associated with this gas cell $i$ (in units of inverse volume), which scales with its distance $r_i$ to the BH. Essentially, this means that the BH smoothing length may vary over time, in order to enclose a similar number of cells $n_{\rm BH,ngb}\pm4$. In TNG50, TNG100 and TNG300 (the TNG simulations with box side lengths of $\sim 50$, $100$ and $300\,\rm cMpc$ respectively), $n_{\rm BH,ngb}$ is set to 512, 256 and 128 \citep[see also Appendix B in][]{Weinberger2017}. In what follows, we refer to the sphere of radius $h_{\rm BH}$ as the BH smoothing volume. In this work, the value of $h_{\rm BH}$ varies from a few hundred $\rm pc$ to a few $\rm kpc$. This parameter increases when the gas density near the BH decreases (as a result of the gas cell width expanding), which preferentially occurs after efficient AGN feedback episodes.

To first order, the rate of inflowing gas mass onto the BH corresponds to the Bondi accretion rate capped at the Eddington limit, that is: 

\begin{align}
    \label{eq:minf}
    &\dot{M}_\mathrm{BH,inf} = \min{\left(\dot{M}_\mathrm{Bondi},\dot{M}_\mathrm{Edd}\right)}
\end{align}

In addition, \citet{Vogelsberger2013} introduced a modification to the BH accretion rate to prevent the formation of a hot bubble of low-density gas around the BH, whenever there is no star-forming gas in its immediate vicinity. As such a configuration would overestimate the BH accretion rate, $\dot{M}_\mathrm{BH,inf}$ is lowered by a factor $(P_{\rm ext}/P_{\rm ref})^2$ whenever $P_{\rm ext}$ (defined as the kernel-weighted pressure of gas in the BH smoothing volume) is lower than a reference pressure $P_{\rm ref}=(\gamma-1)\rho_{\rm sfr}u_{\rm eq}$. In this equation, $\gamma$ is the adiabatic index of the gas, $\rho_{\rm sfr}$ is the gas density threshold beyond which star-formation happens, and $u_{\rm eq}$ is the equilibrium thermal energy when the cooling losses in the BH smoothing volume balance the AGN feedback energy injected. 

When flowing towards the SMBH, gas forms an accretion disk where some fraction of the inflowing mass (with $\epsilon_r=0.2$ in TNG) is converted into radiation, such that the actual BH accretion rate $\dot{M}_\mathrm{BH}$ is:

\begin{align}
    \label{eq:mbh}
    &\dot{M}_\mathrm{BH} = \left(1-\epsilon_r\right)\dot{M}_\mathrm{BH,inf}
\end{align}

Numerically, the SMBH grows continuously according to Eq.~\ref{eq:mbh}, by draining gas mass from the BH smoothing volume cells, in a kernel and volume weighted manner.


\subsubsection{AGN feedback in the radiatively efficient regime}
\label{subsubsec:quasar-tng}

As explained in detail by \citet{Weinberger2017}, AGN feedback in TNG operates in a dual mode, with different AGN energy injection schemes depending on the BH accretion regime. This prescription is motivated by the physical properties of BH accretion disks: when the SMBH accretes at or above a few $10^{-2}$ of $\dot{M}_\mathrm{Edd}$, the AGN is in a radio-quiet, radiative mode, releasing energy from an optically thick and geometrically thin accretion disc \citep{Shakura&Sunyaev1973}. Because this system radiates its heat very efficiently, this is also referred to as the SMBH radiatively efficient regime. Otherwise, the AGN energy is deposited in an optically thin and geometrically thick, radiatively inefficient accretion structure, as described by the Advection-Dominated Accretion Flow model (ADAF, \citealp{Narayan&Yi1994}) or the Adiabatic Inflow-Outflow Solutions (ADIOS, \citealp{Blandford&Begelman1999}). As noted above, super-Eddington accretion is not accounted for in TNG, and we also do not allow it throughout the work presented here.

To determine whether the SMBH is in a radiatively efficient or inefficient state, the common procedure is to compare the Eddington ratio $f_\mathrm{Edd} = \dot{M}_\mathrm{Bondi} / \dot{M}_\mathrm{Edd}$ to a certain value $\chi=10^{-3}-10^{-1}$ \citep[e.g.][]{Sijacki2007,Merloni&Heinz2008,Dubois2012}, as determined from observations of soft-to-hard state transitions of X-ray binaries \citep{Maccarone2003}. Instead of adopting a constant value for $\chi$, the TNG simulations adopt a mass-dependent parametrisation, in order to promote (prevent) the triggering of the radiatively inefficient feedback mode for high-mass (low-mass) BHs, where:

\begin{align}
    \label{eq:chi}
    & \chi = \min\left[0.002\times\left(\frac{\rm M_{\rm BH}}{10^8\, \rm M_\odot}\right)^2,0.1\right]
\end{align}

Whenever $f_\mathrm{Edd} > \chi$, energy is deposited following the TNG thermal model that we also refer to as the \isoth{} model. In this radiatively efficient regime, a fraction $\epsilon_{\rm f}$ of the luminosity $L_{\rm r}$ radiated during the accretion process thermally couples to the gas. In the TNG simulations, $\epsilon_{\rm r}=0.2$, and $\epsilon_{\rm f}=0.1$. Per injection event (during a timestep $\Delta t$), a total energy rate $\dot{E}_{\rm BH}$ is distributed thermally, spherically and isotropically, in a kernel-weighted fashion, to the $n_{\rm BH,ngb}$ gas cells enclosed in the BH smoothing volume, such that:

\begin{align}
    \label{eq:ebh}
    &\dot{E}_{\rm BH}=\epsilon_{\rm f}L_{\rm r}=\epsilon_{\rm f}\epsilon_{\rm r}\dot{M}_{\rm BH,inf}c^2
\end{align}

Therefore, each gas cell $i$ is given a thermal energy increment $\Delta{E}_{i}$, where $m_{\rm tot}$ is the total kernel-weighted gas mass enclosed in the BH smoothing volume:

\begin{align}
    \label{eq:ethcell}
    &\Delta{E}_{i} = \dot{E}_{\rm BH}\,\Delta t\, w_{\rm k}(r_i) \frac{m_{i}}{m_{\rm tot}}
\end{align}

Particularly relevant in the high-accretion regime, when the bolometric luminosity is the highest, \citet{Vogelsberger2013} introduced what they refer to as a radiative, electro-magnetic AGN feedback channel. This channel consists of depositing thermal energy, and is included in the TNG simulations. More specifically, this phenomenological model aims at capturing the effect of the ionising radiation emitted by SMBHs on the thermal state of the gas. For this purpose, the fraction of AGN bolometric luminosity that does not couple to the gas is combined with a redshift-dependent UV background \citep{FaucherGuiguere2009}, which is then used to infer metal line cooling and heating rates from pre-tabulated tables. To determine their net cooling rate, each gas cell is assigned a total bolometric intensity, derived from the obscured bolometric luminosity of all the BHs from the simulation, depending on their distance to the gas cell. This radiative feedback channel renders the TNG thermal feedback more efficient, as star formation is slightly more suppressed in the high-accretion regime compared to when no AGN radiative feedback is included \citep[see Fig. 15 from][]{Vogelsberger2013}.

\subsubsection{AGN feedback in the radiatively inefficient regime}
\label{subsubsec:jet-tng}

When $f_\mathrm{Edd} < \chi$ (defined in Eqn.~\ref{eq:chi}), AGN feedback in TNG consists of momentum kicks to the BH's neighbouring gas cells, meant to mimic the effects of kinetic winds and jets. In this low accretion state, $\epsilon_{\rm f}$\footnote{We use a slightly different parametrization of $\epsilon_{\rm f}$ than the one in \citet{Weinberger2017}, who combine $\epsilon_{\rm r}$ and $\epsilon_{\rm f}$ into a single efficiency parameter $\epsilon_{\rm f,kin}$.} is a variable feedback coupling efficiency defined as:

\begin{align}
    \label{eq:efkin}
    & \epsilon_{\rm f}=\min\left(\frac{\rho}{0.01\rho_{\rm SF}},1\right)
\end{align}

Here, $\rho_{\rm SF}\simeq10^{-23}\,\rm g\,cm^{-3}$ is the density threshold for star formation. This non-constant parametrization of $\epsilon_{\rm f}$ in the radiatively inefficient regime weakens the coupling of the BH energy to the surrounding gas at low density.

After several BH accretion events, when a minimum AGN energy of $10\sigma_{\rm DM}^2m_{\rm tot}\,\rm erg$ has accumulated (where $\sigma_{\rm DM}$ is the 1D dark matter velocity dispersion), a feedback event takes place. The total kinetic feedback energy available $\Delta E_{\rm BH}$ (which corresponds to the sum of $\epsilon_{\rm f}L_{\rm r}$ accumulated over several timesteps) is distributed to all gas cells from the BH smoothing volume, such that each cell $i$ receives a momentum kick $\Delta p_{i}$ in a random direction: 

\begin{align}
    \label{eq:ekincell}
    & \Delta p_{i}=m_{i}\,\sqrt{2\frac{\Delta E_{\rm BH}\, w_{\rm k}(r_i)}{\rho}}
\end{align}

The wind direction is the same for all gas cells per injection event, so that momentum is only conserved on a time-averaged basis. We also refer to this TNG kinetic AGN feedback as the \randomw{} model.

\subsection{The \mistral{} model}
\label{subsec:bh-mistral}

When gas flows from galaxies down to their SMBH, radiation is emitted, predominantly in dense regions close to the BH. There, in the so-called broad-line region (and, to a lesser extent, in the narrow-line region), winds can be driven by line, radiation, and magnetic pressure, carrying mass, momentum and energy that affect their surroundings. To model such AGN-driven winds, \citet{Ostriker2010} and \citet{Choi2012} developed a subgrid AGN feedback scheme that they implemented into the SPH code \gadget{} \citep{Springel2005gadget}. Based on their methodology, we implement the \mistral{} model into the moving-mesh code \arepo{} \citep{Springel2010}, as explained in this section. We note that, in this work, \mistral{} operates at all Eddington ratios, in both the radiatively efficient and inefficient regimes (i.e. we do not use any of the TNG AGN feedback models with \mistral{}). In order to focus on a setup as close as possible to the one adopted by \citet{Choi2012}, we do not include the modification in BH accretion introduced by \citet{Vogelsberger2013}, nor their radiative AGN feedback channel (both of which are used in the TNG simulations). In this work, unlike \citet{Choi2012}, we do not consider the radiative feedback from the X-ray radiation associated with BH accretion.

On its journey towards a SMBH, not all the inflowing gas mass reaches its final destination: instead, some of the gas will be entrained into a wind, outflowing at a rate $\dot{M}_{\rm BH,wind}$. Therefore, the BH accretion rate can be re-written as:

\begin{align}
    \label{eq:mbh-outf}
    & \dot{M}_{\rm BH}=\dot{M}_{\rm BH,inf}-\dot{M}_{\rm BH,wind},
\end{align}

where $\dot{M}_{\rm BH,inf}$ is defined in Eq.~\ref{eq:minf} as the usual Eddington-limited Bondi accretion rate. Eq.~\ref{eq:ew} defines the energy released by the SMBH as $\dot{E}_{\rm BH}$, where $v_{\rm w}$ is the average wind velocity. In this equation, we relate the kinetic energy deposition of the outflowing material to the accretion rate via the AGN feedback efficiency parameter $\epsilon_{\rm w}$, which translates how efficiently the accreted matter turns into wind energy.

\begin{align}
    \label{eq:ew}
    &\dot{E}_{\rm BH}=\epsilon_{\rm w}\dot{M}_{\rm BH}c^2 = \frac{1}{2}\dot{M}_{\rm BH,wind}\,v_{\rm w}^2
\end{align}

Correspondingly, the BH wind momentum rate $\dot{p}_{\rm BH}$ can be defined as:

\begin{align}
    \label{eq:pw}
    &\dot{p}_{\rm BH}=\dot{M}_{\rm BH,wind}v_{\rm w}
\end{align}

Using Eq.~\ref{eq:ew}, we can introduce, as done by \citet{Ostriker2010} and \citet{Choi2012}, a dimensionless parameter $\psi$ such that:

\begin{align}
    \label{eq:psi}
    &\psi = \frac{\dot{M}_{\rm BH,wind}}{\dot{M}_{\rm BH}}=\frac{2\epsilon_{\rm w}c^2}{v_{\rm w}^2}
\end{align}

Using this parameter, mass, energy, and momentum conversion equations can be written as follows:

\begin{align}
    \label{eq:psi-mbh}
    \dot{M}_{\rm BH} &= \frac{1}{1+\psi} \dot{M}_{\rm BH,inf},\\
    \label{eq:psi-mout}
    \dot{M}_{\rm BH,wind} &= \frac{\psi}{1+\psi} \dot{M}_{\rm BH,inf},\\
    \label{eq:psi-ebh}
    \dot{E}_{\rm BH} &= \epsilon_{\rm w}  \frac{1}{1+\psi} \dot{M}_{\rm BH,inf} c^2,\\
    \label{eq:psi-p}
    \dot{p}_{\rm BH} &= \frac{\psi}{1+\psi} \dot{M}_{\rm BH,inf}v_{\rm w}
\end{align}

\begin{figure}
    \centering
    \includegraphics[width=\columnwidth]{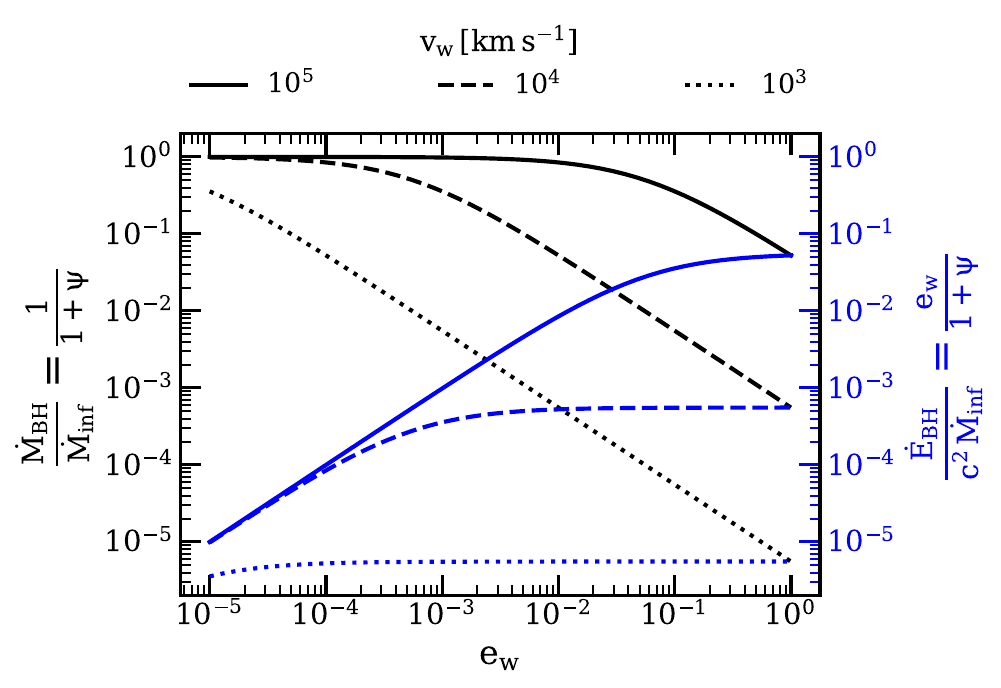}
    \caption{Fraction of the inflowing gas mass rate $\dot{M}_{\rm inf}$ that contributes to the BH mass accretion rate $\dot{M}_{\rm BH}$ (black lines, left y-axis) and to the BH energy rate $\dot{E}_{\rm BH}$ (blue lines, right y-axis) as a function of the AGN wind feedback coupling efficiency $\epsilon_{\rm w}$. Solid, dashed and dotted lines respectively correspond to an average wind velocity of $10^5$, $10^4$ and $10^3\,\rm km\,s^{-1}$.}
    \label{fig:ew-vw}
\end{figure}

Using X-ray and UV absorption line spectroscopy, studies of broad absorption line quasars and ultra-fast outflows suggest that AGN winds velocities can span from thousands to tens of thousands of $\rm km\,s^{-1}$, with inferred kinetic efficiencies estimated to lie between $10^{-4}-10^{-3}$ \citep{Tombesi2011,Tombesi2013,Matzeu2023}. However, these observational constraints remain uncertain, and may vary depending on BH masses and accretion rates. For simplicity, and given the observational uncertainties, this work treats these two parameters as constants, with the aim of capturing the average impact of AGN winds on galaxy evolution at resolved scales. For typical values $\epsilon_{\rm w}=5\times10^{-3}$ and $v_{\rm w}=10^4\,\rm km\,s^{-1}$, $\psi=9$, such that a momentum flux $\dot{p}_{\rm BH}=0.3L_{\rm r}/c$ is injected. This also means that 10\% of the inflowing gas mass rate will be actually accreted onto the BH, driving the remaining 90\% as a (subgrid) wind. More generally, Figure~\ref{fig:ew-vw} shows the fraction of inflowing gas mass accreted by the BH (in black, left y-axis) and the fraction of inflowing rest-mass energy released as a wind (in blue, right y-axis) as a function of $\epsilon_{\rm w}$. The three sets of curves correspond to an average wind velocity of $10^5$, $10^4$ and $10^3\,\rm km\,s^{-1}$ (solid, dashed and dotted lines). By definition, the energy released is lower for low $\epsilon_{\rm w}$ values, allowing the amount of BH mass accreted to be correspondingly higher. In addition, at a fixed $\epsilon_{\rm w}$, a higher wind velocity $v_{\rm w}$ will drive a smaller fraction of the inflowing gas into the outflowing wind, and a higher fraction into actual BH accretion (see Eq.~\ref{eq:ew} and \ref{eq:psi}). Thus, these two parameters control the amount of mass and energy loading in our model, and both the fractions of mass accreted and energy released are more sensitive to $v_{\rm w}$ than to the coupling efficiency $\epsilon_{\rm w}$. 

\begin{figure*}
    \centering
    \includegraphics[width=\textwidth]{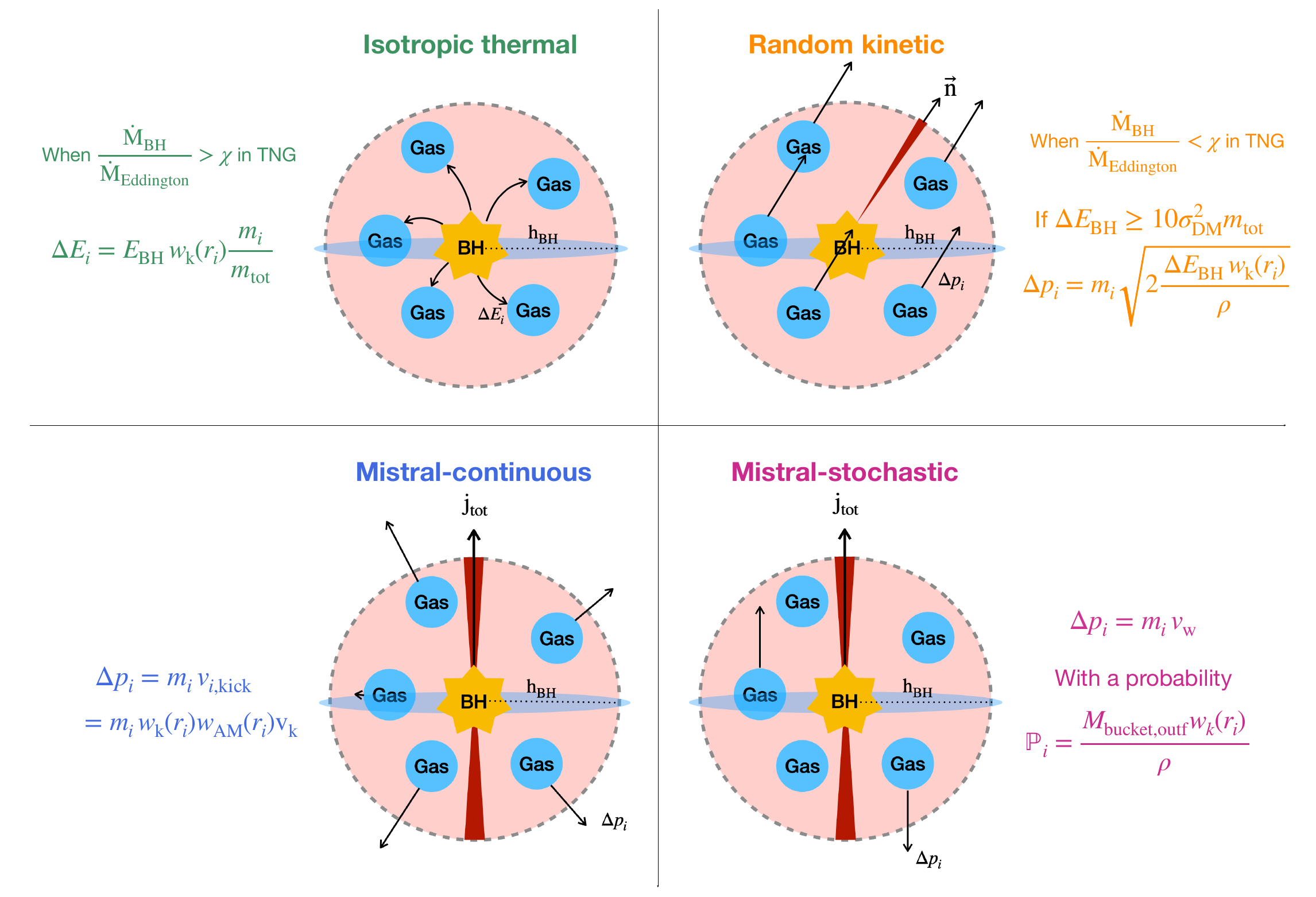}
    \caption{Clockwise: schematic illustration of the \isoth{} and \randomw{} models used in TNG, \mistralS{} and \mistralC{}. For each model, we represent the BH sink particle as a yellow star, surrounded by gas cells depicted by blue circles. The red circle shows the extent of the spherical smoothing volume with radius $h_{\rm BH}$. With the \isoth{} model, each gas cell receives a mass-weighted fraction of the BH energy in a thermal form (Eq.~\ref{eq:ethcell}). With the \randomw{} model, gas cells receive a momentum kick in the same random direction, provided that enough BH energy has accumulated (Eq.~\ref{eq:ekincell}). In the TNG simulations, this model acts when the Eddington ratio is below a certain threshold $\chi$ (Eq.~\ref{eq:chi}), and the \isoth{} model operates otherwise. With \mistralC{}, cells are kicked radially away from the BH, with a velocity weighted by the alignment of the cell to the angular momentum of the gas surrounding the BH (Eq.~\ref{eq:vkick}). With \mistralS{}, cells are kicked at a velocity $v_{\rm w}$ with a probability $\mathbb{P}$ (Eq.~\ref{eq:effproba-outf}), in a direction parallel or anti-parallel to the gas angular momentum.} 
    \label{fig:schematic}
\end{figure*}

In this paper, the BH inflowing gas mass rate is determined following the Bondi prescription. The total gas mass accreted by the BH during a given timestep $\Delta t$ follows Eq.~\ref{eq:psi-mbh}, and this gas mass is continuously swallowed from gas cells that belong to the BH smoothing volume, as described in Sec.~\ref{subsubsec:quasar-tng}. Once gas is drained, the BH energy resulting from the accretion process is distributed over the remaining gas mass, at the same timestep. This is done in a kinetic fashion, either in a continuous way with \mistralC{}, or in a stochastic manner with \mistralS{}. \mistralC{} provides a kinetic alternative to the TNG thermal model, as both models release AGN feedback continuously in all the cells from the BH smoothing volume. Conversely, \mistralS{} inputs momentum only in a subset of stochastically selected cells. While \mistralS{} is a closer analogue to the feedback model developed by \citet{Ostriker2010} and \citet{Choi2012}, \mistralC{} will be used to illustrate the transition from 1) a purely thermal to a kinetic, radiatively efficient AGN feedback model and 2) a continuous to a stochastic feedback injection scheme. We now review the details of these two flavours of the \mistral{} model, which are tested and used separately in this study. Both versions of \mistral{} are schematically illustrated and compared to the TNG AGN feedback models in Fig.~\ref{fig:schematic}.

\subsubsection{\mistralC{}: continuous radial momentum injection}
\label{subsec:mistral1}

In \mistralC{}, gas cells are kicked radially away from the BH, at a velocity $v_{\rm kick}$, such that both momentum and energy are conserved. For this purpose, we adopt the following procedure. We first adjust the direction of the kicks by subtracting the center of mass to each radial vector $ \bmath{r_i}$, so that each cell $i$ is kicked following the unit vector $\bmath{n_i}=\bmath{r_i}/||\bmath{r_i}||-\sum_i m_i w_{\rm k}(r_i) \bmath{r_i}/\sum_i m_i w_{\rm k}(r_i)$. To favour the emergence of a bipolar outflow out of this spherically symmetric injection scheme, we define $v_{\rm kick}$ such that a higher velocity increment is imparted to gas cells that are aligned with the mass-weighted angular momentum $\bmath{j_{\rm tot}}$ of the smoothing volume (close to the BH):

\begin{align}
    \label{eq:vkick}
    &\Delta p_i = m_i \,v_{i,\rm kick}=m_i\, w_{\rm k}(r_i) w_{\rm AM}(r_i) \mathrm{v_{kick}}
\end{align}

Here, $w_{\rm AM}(r_i) = ||\bmath{r_i}\cdot\bmath{j_{\rm tot}}||$ corresponds to an angular momentum weight, and $\mathrm{v_{kick}}$ is defined to ensure energy conservation such that:

\begin{align}
    \label{eq:econs}
    &E_{\rm BH} = \sum_i \frac{\left(\bmath{p_i}+\bmath{\Delta p_i}\right)^2}{2 m_i}  - \frac{\bmath{p_i}^2}{2 m_i},
\end{align}

Where, $\bmath{p_i}=m_i\,v_i$ is the current momentum of cell $i$ (before the injection of AGN feedback). Solving Eq.~\ref{eq:econs} and determining the value of $\mathrm{v_{kick}}$ is equivalent to solving the quadratic equation $a_{\rm tot} \mathrm{v_{kick}}^2 + b_{\rm tot} \mathrm{v_{kick}} - {E}_{\rm BH} = 0$, which gives:



\begin{align}
    \label{eq:fvel}
    &\mathrm{v_{kick}} = \frac{b_{\rm tot}+\sqrt{b_{\rm tot}^2+4 E_{\rm BH} a_{\rm tot}}}{2 a_{\rm tot}},\\
    &a_{\rm tot} = \sum_i \frac{m_i\left[w_{\rm k}(r_i) w_{\rm AM}(r_i)\bmath{n_i}\right]^2}{2},\\
    &b_{\rm tot} = \sum_i \frac{m_iw_{\rm k}(r_i) w_{\rm AM}(r_i)\bmath{n_i}\cdot\bmath{v_i}}{2}
\end{align}

Once determined, $\mathrm{v_{kick}}$ is then used to update each cell velocity, momentum and energy. While this prescription ensures conservation of energy, it does not strictly distribute the BH energy to the gas mass expected from Eq.~\ref{eq:psi-mout}. Indeed, we implicitly assume that all gas mass in the BH vicinity which is not accreted will be kicked, at a certain velocity determined such that we distribute the total AGN energy. As a consequence, if there is a large number of cells over which we must distribute a low BH energy, the effective velocity $v_{i, \rm kick}$ imparted to a cell (corresponding to the the wind velocity at the scale of $h_{\rm BH}$) may be orders of magnitude lower than the input wind velocity $v_{\rm w}$ (corresponding to the wind velocity at the accretion scale). In order to inject the correct amount of momentum, as determined by Eq.~\ref{eq:pw} and \ref{eq:psi-p}, when Eq.~\ref{eq:psi-mout} is not fulfilled, one solution might be to change the amount of gas mass over which momentum is injected (e.g. by changing the size of the BH smoothing volume), which we defer for future work\footnote{An alternative approach to conserve both the total energy and momentum exactly (regardless of the gas mass involved in the feedback event) would be to inject momentum via kinetic kicks, and then distribute any residual energy as thermal energy. However, in cases where the gas mass receiving feedback is significantly larger than the target wind mass, this hybrid scheme would result in feedback being dominated by thermal energy. Since our goal with \mistralC{} is to explicitly investigate the impact of kinetic feedback, in contrast to more common thermal implementations in the radiatively efficient regime, we deliberately choose to preserve energy at the expense of exact momentum injection.}. In this paper, to ensure a meaningful comparison with the TNG feedback models, we instead choose to distribute the BH energy to all gas cells within the BH smoothing volume, even though the momentum injection then differs from what is specified by Eq.~\ref{eq:pw} and \ref{eq:psi-p}.

\subsubsection{\mistralS{}: stochastic bipolar momentum injection}
\label{subsec:mistral2}

\mistralS{} aims at being the \arepo{} equivalent of the AGN wind model from \citet{Ostriker2010} and \citet{Choi2012}. Instead of continuously injecting momentum, we now switch to a stochastic approach: gas cells from the BH smoothing volume are kicked at the chosen input velocity $v_{\rm w}$, with a probability $\mathbb{P}_i$ for each cell $i$:

\begin{align}
    \label{eq:proba-outf}
    &\mathbb{P}_i=\frac{\dot{M}_{\rm BH,wind}\Delta t w_k(r_i)}{\rho}
\end{align}

This ensures that we will impart momentum only to the fraction of $\dot{M}_{\rm BH,inf}$ that should be launched in a wind, as defined in Eq.~\ref{eq:psi-mout}. Each cell eligible for receiving momentum is then kicked with a velocity $v_{\rm w}$ in a direction parallel (anti-parallel) to the mass-weighted angular momentum of the gas $j_{\rm tot}$ of the BH smoothing volume, if the gas cell is above (below) the midplane perpendicular to $j_{\rm tot}$\footnote{We checked that randomly kicking the cells parallel or anti-parallel to $j_{\rm tot}$ (as opposed to imposing the direction of the velocity kick based on the position of the cell relative to the galaxy disk) barely changes the impact of the \mistralS{} model on galaxy and BH properties.}.


Because we cannot kick a fraction of the gas mass contained in a cell (this would require changing the refinement strategy), we allow an AGN feedback event only if $\dot{M}_{\rm BH,wind}\Delta t$ is larger than the gas mass resolution. Therefore, at a given timestep, the probability $\mathbb{P}_i$ requires a sufficiently high BH accretion rate to have a chance to be satisfied. To circumvent this issue, we redefine the effective probability for gas cell $i$ to be outflowing as:

\begin{align}
    \label{eq:effproba-outf}
    &\mathbb{P}_i=\frac{M_{\rm bucket,outf} w_k(r_i)}{\rho}
\end{align}

Where $M_{\rm bucket,outf}=\sum_j\left(\dot{M}_{\rm BH,outf,j}\Delta t_j\right)$ is a bucket variable used to keep track of the outflowing gas mass expected after each timestep $\Delta t_{\rm j}$. When a gas cell is kicked, we remove its mass from $M_{\rm bucket,outf}$ at the end of the timestep. Therefore, the probability of receiving a velocity increment at a given timestep remains the same for all the eligible gas cells. In order to ensure that we effectively kick the outflowing gas mass expected on a time-average basis, gas cells are eligible for being kicked only if their mass is lower than the $M_{\rm bucket,outf}$. We eventually update the velocity, momentum and internal energy of the cells that have been kicked. Given the stochastic nature of this model, the injected energy and momentum correspond, on average over several AGN feedback events, to the values from Eq.~\ref{eq:psi-ebh} and \ref{eq:psi-p}, and typically differ by no more than 20 per cent per injection event.

\section{Idealised Milky Way-mass simulations}
\label{section:resultsm12}

A major goal of this work is comparing the impact of \mistral{} on galaxy properties, compared to the AGN feedback models used in the TNG simulations. For this purpose, we first use idealised Milky Way-mass simulations in this section, before applying a similar analysis on $z=2$ cosmological zoom simulations (Sec.~\ref{section:resultsz2}).

\subsection{Setup of the idealised simulations}
\label{subsec:gal-ics}

To examine the impact of \mistral{} winds on galaxies, we first use simulations of an idealised Milky Way-mass disc-dominated galaxy as a controlled experiment. Specifically, the initial conditions correspond to those of galaxy m12 from \citet{Su2019}, converted from \gizmo{} \citep{Hopkins2015} to \arepo{}-readable data. We summarise some of the galaxy properties below and in Table~\ref{tab:run_prop}, and refer the reader to Section 2.2 of \citet{Su2019} for a more complete description.

The galaxy disc is embedded in a box of $3000\,\rm kpc$ in width, and is hosted in a $1.5\times10^{12}\,\rm M_\odot$ dark matter halo, which follows a spherical, isotropic NFW density profile \citep{NFW1996}. The dark matter halo has a virial radius\footnote{We define the virial radius as the radius of the sphere that encloses a mean density that is 200 times the critical density of the Universe.} of $R_{\rm vir}=230\,\rm kpc$, a NFW scale length of $20.4\,\rm kpc$, a concentration parameter $c=12$, and a maximum circular velocity of $174\,\rm km\,s^{-1}$. The galaxy contains a stellar bulge that follows a \citet{Hernquist1990} profile, and exponential, rotation-supported gas and stellar disks, in which the initial temperature is at pressure equilibrium. In order to roughly reproduce the observed Milky Way profile \citep{Miller&Bregman2013,Miller&Bregman2015}, the initial conditions include a hydrostatic, spherical gaseous halo with a mass of $1.5\times10^{11}\,\rm M_\odot$, following a beta profile with a scale radius of $20.4\,\rm kpc$ and $\beta=0.5$.

Dark matter is modelled by collisionless particles with a mass of $m_{\rm DM}=5\times10^5\,\rm M_\odot$. The baryon mass resolution (for both gas and stellar particles) has a mean value of $m_{\rm baryons}=8\times10^4\,\rm M_\odot$, and can vary by a factor of two for gas upon cell (de)refinement or by mass losses for the stellar particles. The Plummer equivalent gravitational softening of the collisionless component (dark matter, stellar and wind particles) is fixed to $\rm \epsilon_{DM,stars}=195\,pc$, and the gravitational softening of the gas is set to 2.5 times the cell radius, with a minimum value of $\rm \epsilon_{gas}=50\,pc$.

The galaxy includes exponential gas and stellar disks, with a scale length of 6 and 3 kpc, respectively, and a scale height of 0.3 kpc for both. The galaxy also features an extended spherical gas halo with a scale radius of 20 kpc. In what follows, we define the galaxy disk as a cylinder of $R_{\rm disk} = 20\,\rm kpc$ in radius and $h_{\rm disk} = 2\,\rm kpc$ in height. Initially, this disk has a gas fraction $f_{\rm gas}=M_{\rm gas}/(M_{\rm gas}+M_{\rm star})\simeq 7\%$. It contains a gas mass $M_{\rm gas}=4.2\times10^{9}\,\rm M_\odot$, hosts a stellar mass $M_{\rm star}=5.7\times10^{10}\,\rm M_\odot$, and has a metallicity $Z=Z_\odot=0.02$ at the core. Gas metallicity decreases as a function of radius $r$ as $Z_\odot\times(0.05+0.95/(1+(r/R_{\rm disk})^{1.5}))$, until it reaches $Z=0.001$ outside the disk. We first run the initial conditions adiabatically, without feedback and cooling processes, for 100 Myr, in order to relax the \arepo{} mesh before turning on star formation and stellar feedback for another 500 Myr. These two steps respectively allow the \arepo{} grid and the galaxy disk to stabilise, so that the galaxy reaches an equilibrium state, before we allow the central SMBH to accrete gas and release energy. Throughout this section, we refer to $t=0\,\rm Myr$ as the time at which BH physics is enabled. During the 500 Myr pre-BH phase, a stellar mass of $9\times10^{9}\,\rm M_\odot$ forms, so that at $t=0\,\rm Myr$, the galaxy disk has a gas mass $M_{\rm gas}=3.3\times10^{9}\,\rm M_\odot$, a total stellar mass $M_{\rm star}=6.6\times10^{10}\,\rm M_\odot$ and a gas fraction $f_{\rm gas}=0.05$.

The SMBH is initially positioned at the center of the simulation box, and at the center of the galaxy disk. The SMBH has an initial mass of $M_{\rm BH}=5\times10^7\,\rm M_\odot$, lying between the local BH mass-stellar mass scaling relations from \citet{Kormendy&Oh2013} and \citet{Reines&Volonteri2015}. Although this BH seed mass is an order of magnitude greater than the SMBH in the Milky Way, it enables the BH to grow without requiring an accretion boost parameter -- which  would otherwise be necessary to examine the effects of different feedback prescriptions, given the short duration and idealised nature of these simulations. The BH grows via gas accretion and releases energy as described in Section~\ref{section:models}, depending on the AGN feedback model used. Specifically, in this section, we study the following set of simulations: 

\begin{itemize}
    \item \nobh{}: without BH accretion and feedback.
    \item \isoth{}: isotropic thermal BH energy release, corresponding to the thermal feedback used in TNG at high Eddington ratios (in the radiatively efficient regime). We set $\epsilon_{\rm r}=0.1$ and $\epsilon_{\rm f}=5\times10^{-3}$, and use this feedback model at any Eddington ratio. Note that here, $\epsilon_{\rm r} \times \epsilon_{\rm f}=5\times10^{-4}$, unlike 0.02 in the TNG simulations.
    \item \randomw{}: random AGN winds, corresponding to the AGN kinetic feedback used in TNG at low Eddington ratios (in the radiatively inefficient regime). Similarly as for the \isoth{} simulation, we use $\epsilon_{\rm r}=0.1$ and $\epsilon_{\rm f}=5\times10^{-3}$, and we use the \randomw{} model at all Eddington ratios (unlike what is done in the fiducial TNG simulations).
    \item \tng{}: default TNG feedback models and parameters. In the radiatively efficient and inefficient regimes, $\epsilon_{\rm f}=0.1$ or follows Eq.~\ref{eq:efkin}, respectively, and $\epsilon_{\rm r}=0.2$. In practice, the SMBH never enters the radiatively inefficient regime during the simulation runtime, such that $\epsilon_{\rm r} \times \epsilon_{\rm f}=0.02$ at all times. This simulation is the only one of the set that includes the radiative AGN feedback developed by \citet{Vogelsberger2013}.
    \item \mistralC{}: continuous radial momentum injection mode of \mistral{}. We set $\epsilon_{\rm w}=5\times10^{-4}$ and $v_{\rm w}=10^4\,\rm km\,s^{-1}$.
    \item \mistralS{}: stochastic bipolar momentum injection mode of \mistral{}. We set $\epsilon_{\rm w}=5\times10^{-4}$ and $v_{\rm w}=10^4\,\rm km\,s^{-1}$.
\end{itemize}

In all simulations that include BH physics, we adopt $n_{\rm BH,ngb}=512$ as the weighted number of gas cells in the BH smoothing volume. Except for the simulation with the \tng{} model, we use the same feedback coupling efficiencies in order to better determine how injecting a given BH rest-mass accreted energy using different schemes impacts galaxy properties. We also checked (but do not show) that adopting the same net efficiency for all simulations (i.e. $\dot{E}_{\rm BH}/\dot{M}_{\rm BH,inf}c^2=2.6\times10^{-4}$) does not impact the conclusions drawn in this section. Given our choice of parameters, 52 percent of the inflowing gas mass is accreted in the simulations with \mistral{}, which goes up to 90 percent in the simulations with the \isoth{} and \randomw{} models, and to 80 percent in the simulation with the \tng{} setup. We note that our choice of $v_{\rm w}=10^4\,\rm km\,s^{-1}$ is consistent with the velocity of AGN-driven winds measured at small scales, such as the broad line region \citep[e.g.][]{Tombesi2011, Matzeu2023} and the inner edge of the dusty torus \citep{Choi2020}. This velocity however differs from that of the outflow measured at kpc scales, which may reach lower values \citep[e.g.][]{Choi2022}. We also emphasize that in this section, our aim is to illustrate how the continuous or stochastic injection of thermal or kinetic energy affects galaxy properties. We note that the results from the \tng{} idealised simulation that use the same coupling efficiency parameters as TNG are presented solely for reference. We do not apply these parameters to \mistral{} since doing so would result in 97 per cent of the inflowing mass being expelled as a wind, leaving only a small amount available for black hole accretion. Instead, we adopt a similar parametrisation as that used by \citet{Choi2012} in a similar idealised context, with a wind efficiency within the range of observational estimates. 

\begin{table}
	\centering
	\caption{Initial conditions and properties of the idealised galaxy simulations. \textbf{Left:} $M_{\rm halo}$: dark matter halo mass, $M_{\rm disc}$: baryonic disc mass (gas + stars), $M_{\rm BH,seed}$: BH seed mass, $R_{\rm vir}$: virial radius, $f_{\rm gas}$: gas disc fraction, $Z_{\rm disc}$: disc metallicity. \textbf{Right:} $L_{\rm box}$: length of the simulated box, $m_{\rm DM}$: dark matter mass resolution, $m_{\rm baryons}$: mean baryon mass resolution, $\rm \epsilon_{DM,stars}$: Plummer equivalent gravitational softening of the collisionless component, $\rm \epsilon_{gas}$: minimum value of the gravitational softening of the gas, $n_{\rm BH,ngb}$: weighted number of gas cells in the BH smoothing volume}
	\label{tab:run_prop}
	\begin{tabular}{llcll}
		\hline
        \multicolumn{2}{c}{Galaxy properties} & \vline & \multicolumn{2}{c}{Simulation properties}\\
        \hline
		$M_{\rm halo}$ & $1.5\times10^{12}\,\rm M_{\odot}$ & \vline & $L_{\rm box}$ & $3000\,\rm kpc$\\
        $M_{\rm disc}$ & $6.1\times10^{10}\,\rm M_{\odot}$ & \vline & $m_{\rm DM}$ & $5\times10^5\,\rm M_{\odot}$\\
        $M_{\rm BH,seed}$ & $5\times10^7\,\rm M_{\odot}$ & \vline & $m_{\rm baryons}$ & $8\times10^4\,\rm M_{\odot}$\\
        $R_{\rm vir}$ & $230\,\rm kpc$ & \vline & $\rm \epsilon_{DM,stars}$ & $195\,\rm pc$\\
        $f_{\rm gas}$ & 0.07 & \vline & $\rm \epsilon_{gas}$ & $50\,\rm pc$\\
        $Z$ & $0.05-1\,\rm Z_{\odot}$ & \vline & $n_{\rm BH,ngb}$ & $512$\\
        \hline
	\end{tabular}
\end{table}

\subsection{Comparison of \mistral{} to the TNG AGN feedback models}

Throughout this section, we explore the impact of \mistral{} on gas properties (Sec.~\ref{subsec:m12props}), star formation and BH growth (Sec.~\ref{subsec:m12_sf-bh}), and outflow properties (Sec.~\ref{subsec:m12outf}), compared to the \isoth{} and \randomw{} models used in TNG.

\subsubsection{Effect of AGN feedback on gas properties}
\label{subsec:m12props}

\begin{figure*}
    \centering
    \includegraphics[width=0.92\textwidth]{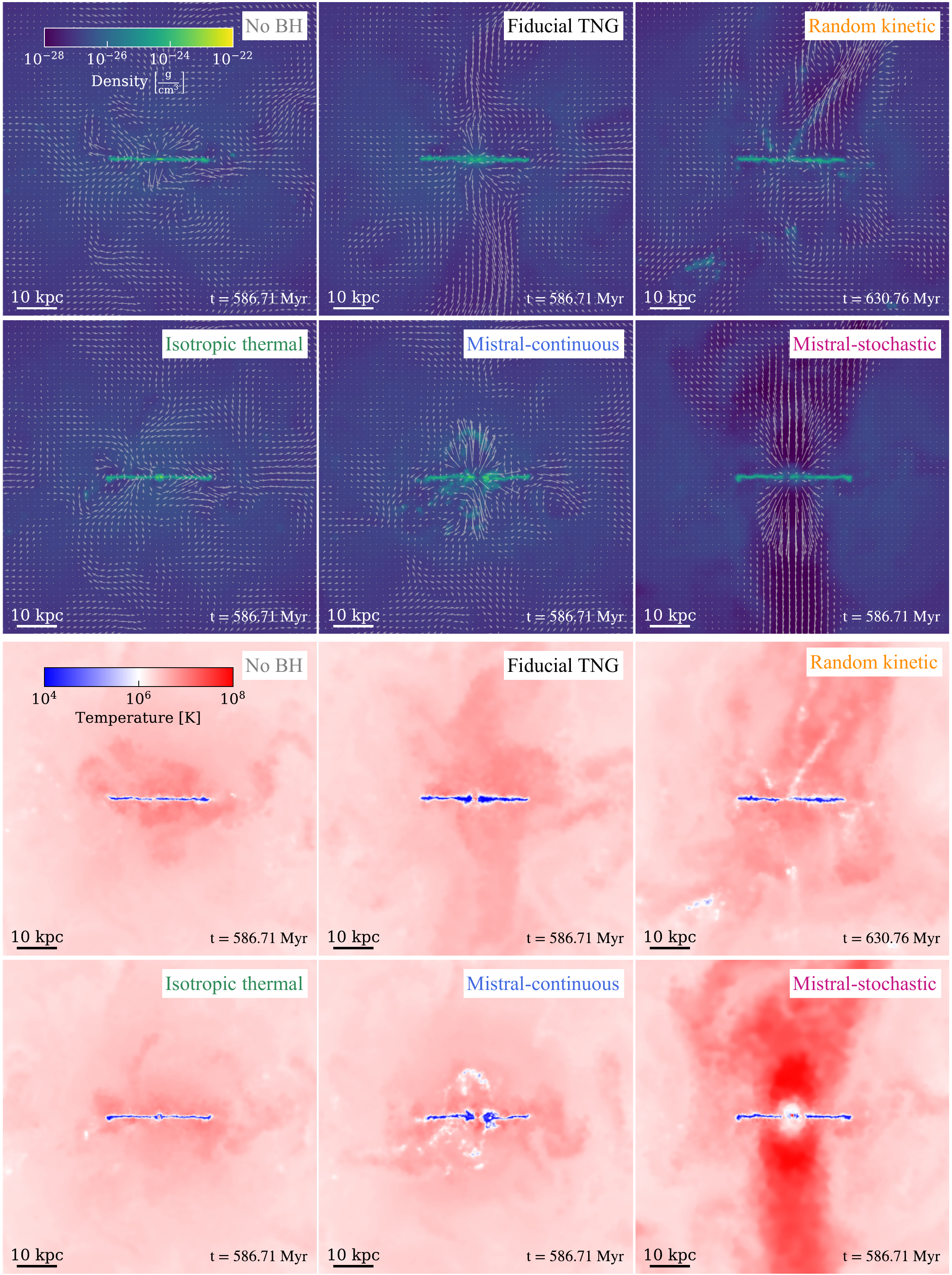}
    \caption{Edge-on 80 kpc wide slices of the gas density (top rows) and temperature (bottom rows). From left to right and top to bottom, we show the \nobh{}, \tng{}, \randomw{}, \isoth{}, \mistralC{} and \mistralS{} simulations. A 10 kpc width scale bar is plotted in the lower left corner, and the time at which these maps are plotted is indicated in the lower right corner. In the density maps, grey arrows show the gas velocity field overlaid.}
    \label{fig:MWmapsall}
\end{figure*}

We begin with a qualitative comparison of the effects of AGN feedback on gas properties. Fig.~\ref{fig:MWmapsall} shows edge-on maps of gas density and temperature, for the simulations without BH physics (\nobh{}), and with the \tng{}, \randomw{}, \isoth{}, \mistralC{} and \mistralS{} AGN feedback models. The maps are plotted at a time $t\simeq586\,\rm Myr$, except in the simulation with the \randomw{} model, for which we show them just after an AGN feedback event occurs, at $t\simeq630\,\rm Myr$.

In the absence of AGN feedback and with the \isoth{} model, gas cycles around the galaxy and the BH without producing a galaxy-scale outflow. Although the \isoth{} model is also at play in the simulation with the \tng{} physics, some gas escapes the galaxy perpendicular to its disk, following the path of least resistance, which creates a thin bipolar wind. In this simulation, we find that the \randomw{} model is never activated, as the Eddington ratio is always between 2 and 10 times higher than the thermal-to-kinetic mode threshold $\chi$ (Eq.~\ref{eq:chi}). Therefore, the difference in gas distribution between the \tng{} and the \isoth{} simulations is the consequence of the TNG thermal feedback being stronger in the former, for which the fraction of BH accreted rest-mass energy coupling to the gas $\epsilon_{\rm f}$ is 20 times higher. With the \randomw{} model, a large scale, one-sided wind is launched in a random direction. Conversely, \mistralC{} generates a bipolar wind whose shock front, composed of dense cold gas, extends up to $10\,\rm kpc$ before flowing back to the ISM at later times. On the other hand, the episodic BH energy injection with \mistralS{} efficiently carves low-density, hot bipolar outflows into the CGM. In these idealised simulations, the angular momentum of gas in the BH vicinity aligns with the z-axis. For this reason, and by construction, gas cells with \mistralS{} are always kicked vertically. As a result, this creates a collimated region depleted of gas after successive AGN feedback episodes. The biconical winds produced with the two versions of \mistral{} are also visible from the velocity field overlaid in grey arrows in the gas density maps.

\begin{figure}
    \centering
    \includegraphics[width=\columnwidth]{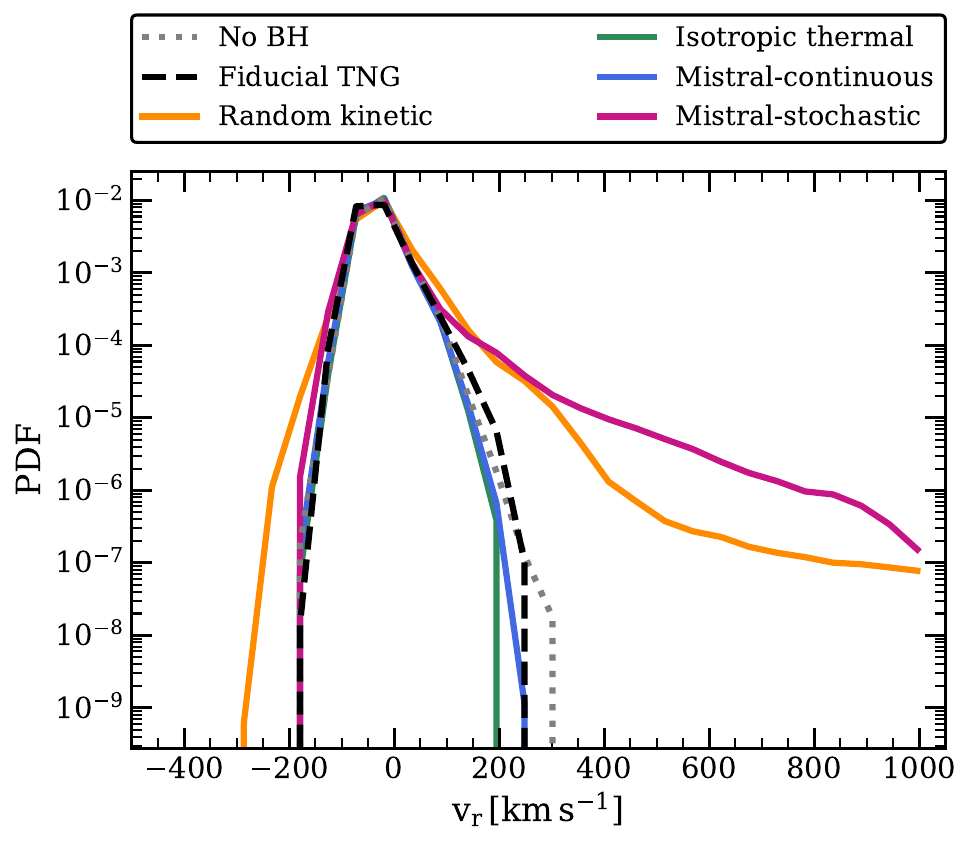}
    \caption{Probability density function of the gas radial velocity with data stacked from 500 to 600 Myr. We restrict the material used in the analysis to gas within half the galaxy virial radius and outside the galaxy disc. We show simulations without BH physics (\nobh{}) in grey dotted, and with the \tng{}, \randomw{}, \isoth{}, \mistralC{} and \mistralS{} feedback models in dashed black, solid orange, green, blue and purple, respectively.}
    \label{fig:vrpdf}
\end{figure}

The gas velocity is further quantified in Fig.~\ref{fig:vrpdf}, which shows the mass-weighted probability density function of radial velocities stacked between $t=500$ to $600\,\rm Myr$ (which encloses the time for which the maps in Fig.~\ref{fig:MWmapsall} are displayed). We plot gas within half the virial radius of the galaxy (i.e. within $\rm 115\, kpc$ from the galaxy center) and that is outside of the galaxy disk (as defined in Sec.~\ref{subsec:gal-ics}), in order to exclude turbulent gas from the ISM and better focus on inflowing and outflowing gas in the CGM. The leftmost part of Fig.~\ref{fig:vrpdf} shows similar velocities of $v_{\rm r}\simeq-200\,\rm km\,s^{-1}$ for the inflowing gas (i.e. negative radial velocities) with all models, with the exception of the \randomw{} model which produces inflows as fast as $v_{\rm r}\simeq-300\,\rm km\,s^{-1}$. However, some models clearly lead to different positive radial velocities. Outside of the galaxy disk in the \nobh{}, \tng{}, \isoth{} and \mistralC{} simulations, gas reaches a maximum radial velocity of $200-300\,\rm km\,s^{-1}$. On the other hand, the \randomw{} model and \mistralS{} are more efficient at generating winds as fast as $1000\,\rm km\,s^{-1}$, and \mistralS{} produces more gas at $v_{\rm r}>300\,\rm km\,s^{-1}$ than the \randomw{} model. Therefore, this shows that injecting kinetic energy stochastically, rather than in a continuous manner or in thermal form, is more efficient at producing fast AGN-driven winds out of the ISM with our setup. This result is similar to what is found for instance by \citet{Choi2012}, using the mechanical feedback model that inspired \mistralS{}.

\begin{figure}
    \centering
    \includegraphics[width=\columnwidth]{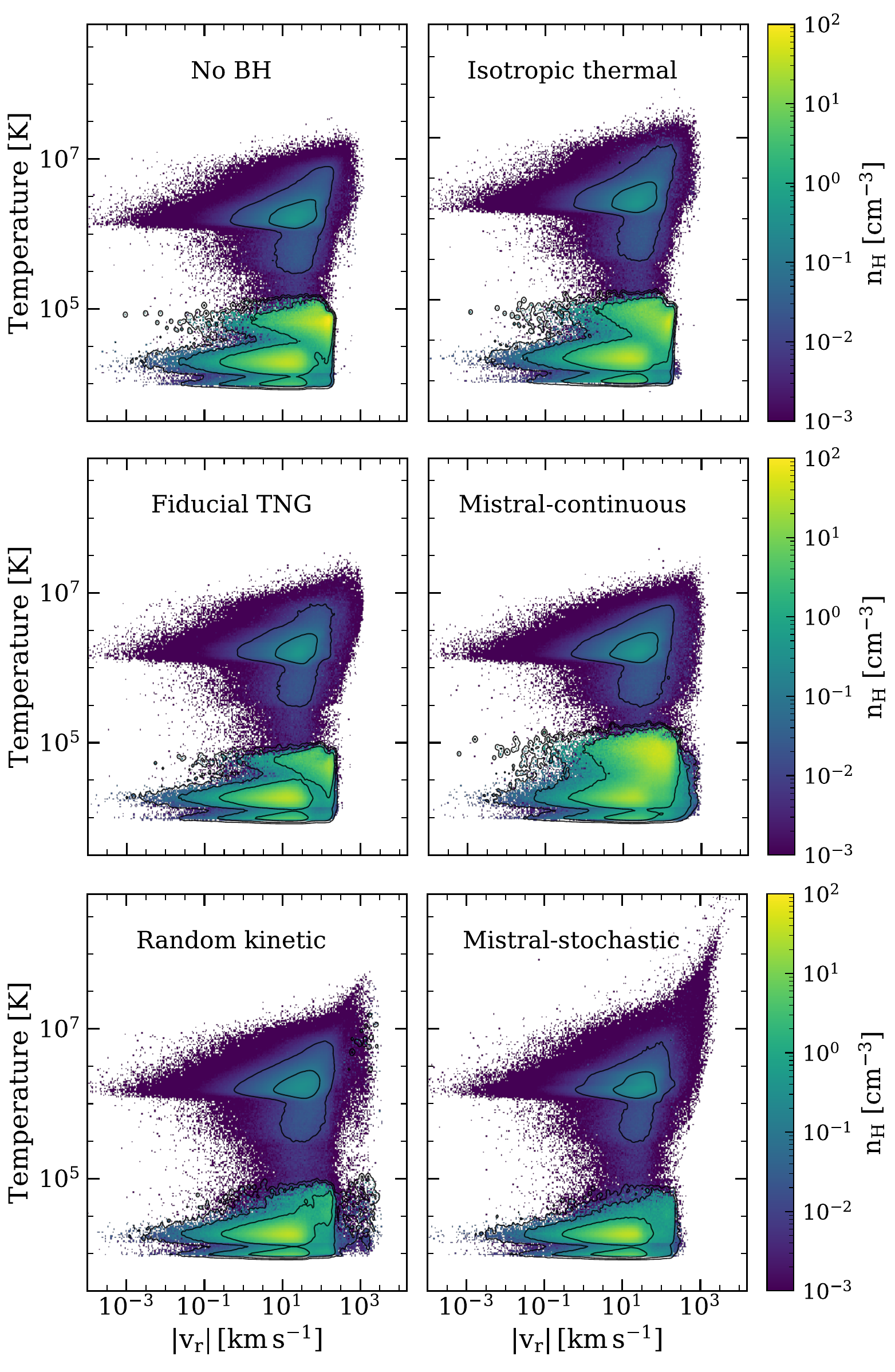}
    \caption{Phase diagrams of gas temperature against the gas radial velocity, color-coded by the mean hydrogen number density. From top to bottom and left to right, we show the \nobh{}, \isoth{} \tng{} \mistralC{}, \randomw{} and \mistralS{} simulations, with data stacked from 500 to 600 Myr (21 snapshots). Black lines show the $1\sigma$ contours for $n_{\rm H}=10^{-2},10^{-1}$ and $1\,\rm cm^{-3}$.}
    \label{fig:phase_diag}
\end{figure}

Fig.~\ref{fig:phase_diag} quantitatively compares the effect of the different AGN feedback models on gas properties. The figure shows phase diagrams of temperature as a function of radial velocity for gas within half the virial radius of the galaxy. We stack data between 500 and 600 $\rm Myr$ to smooth any transient effects, which may arise from the burstiness of star formation and from the stochasticity of feedback processes. We also highlight the $1\sigma$ contours (black lines) where the gas hydrogen density is $n_{\rm H}=10^{-2},10^{-1}$ and $1\,\rm cm^{-3}$, to guide the comparison between the simulations. These contours globally reveal two regimes: dense, relatively cold gas with $T\leq10^5\,\rm K$ which corresponds to ISM gas, and diffuse gas at $T>10^5\,\rm K$, which is predominantly located in the CGM. Because of the two-phase ISM model adopted in these simulations, the temperature of the dense star-forming gas is limited to $10^4\,\rm K$, and can reach higher temperatures as a consequence of over-pressurisation \citep{Springel&Hernquist2003}. Overall the \nobh{}, \isoth{} and \tng{} simulations have similar phase diagrams. While the six galaxies all have dense gas at $T\simeq10^4\,\rm K$, only the simulations with the \randomw{} model and \mistralS{} lack a second reservoir of dense gas at $n_{\rm H}=1\,\rm cm^{-3}$ and $T\simeq10^5\,\rm K$. This gas phase, present in the four other simulations, corresponds to heated and ejected gas, which has formed a small-scale galactic fountain that quickly cycles back to the ISM. With the \randomw{} model and \mistralS{}, this gas is sufficiently accelerated that it is ejected out of the ISM. With \mistralS{} in particular, this produces an extended tail of fast, hot and diffuse gas in the upper right corner of the plot. The phase diagrams also show that \mistral{} accelerates gas to higher velocities than the \isoth{} model. The higher velocity gas is cold and dense with \mistralC{}, and mostly belongs to the galaxy disc (as it is not visible in Fig.~\ref{fig:vrpdf}), while it is predominantly hot and diffuse with \mistralS{}, supporting the qualitative results from Fig.~\ref{fig:MWmapsall}.

\subsubsection{Regulation of star formation and BH growth}
\label{subsec:m12_sf-bh}

\begin{figure}
    \centering
    \includegraphics[width=\columnwidth]{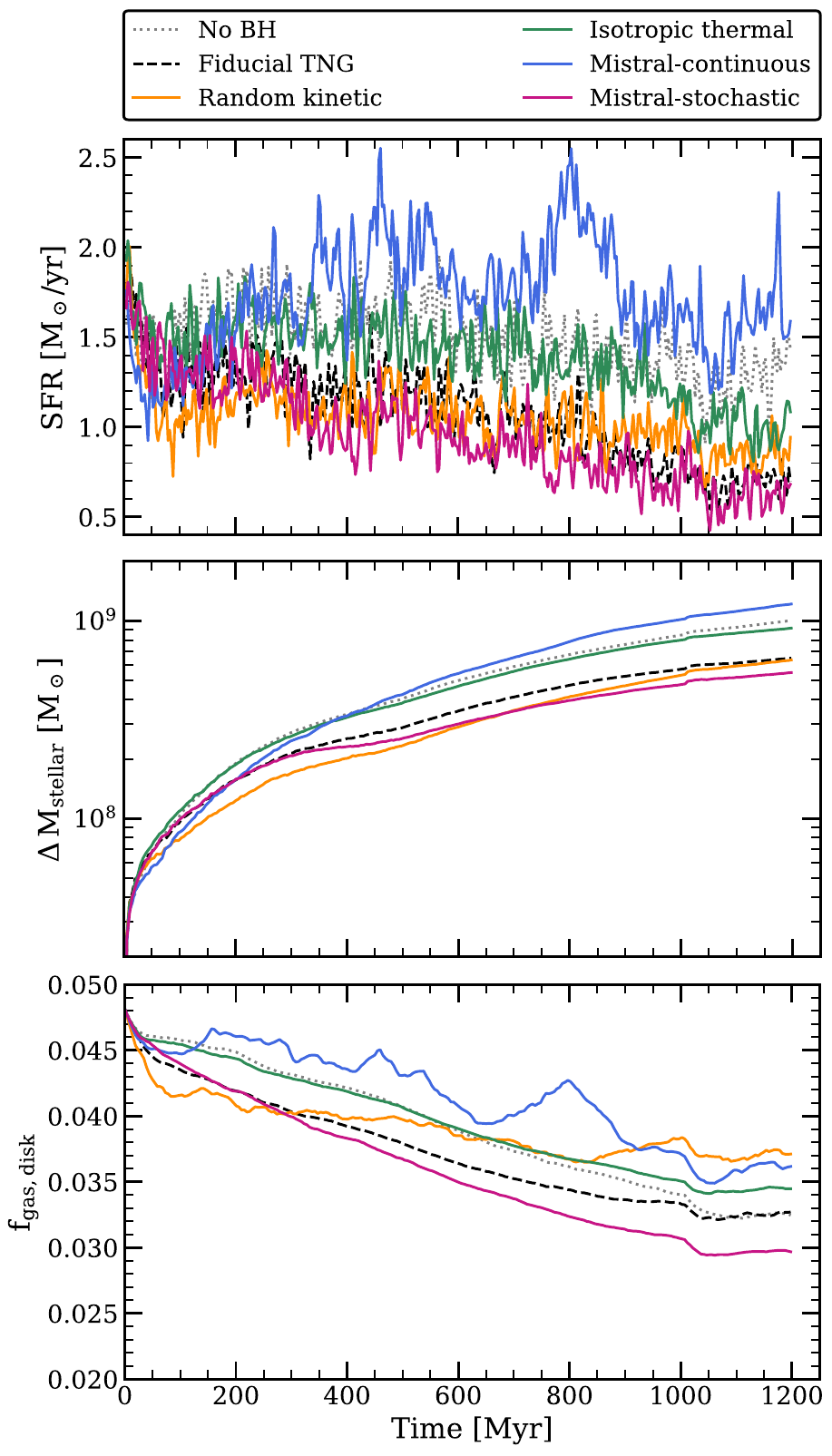}
    \caption{From top to bottom: SFR, cumulative stellar mass formed and gas fraction in the galaxy disc, as a function of time. We show simulations without BH physics in dotted grey, with the \tng{} model in dashed black, and with \randomw{}, \isoth{}, \mistralC{} and \mistralS{} shown by orange, green, blue and purple solid lines, respectively.}
    \label{fig:m12sf_props}
\end{figure}

We now quantify the impact of the different AGN feedback models on regulating star formation and black hole growth. From top to bottom, Fig.~\ref{fig:m12sf_props} shows the star formation rate, cumulative stellar mass formed, and gas fraction in the galaxy disk as a function of time. Here, we define gas fraction as $f_{\rm gas,disk}=M_{\rm gas,disk}/(M_{\rm gas,disk}+M_{\rm stellar,disk})$, where $M_{\rm stellar,disk}$ includes both newly formed stars and the stellar mass formed during the pre-BH phase of the simulations.

\begin{figure}
    \centering
    \includegraphics[width=\columnwidth]{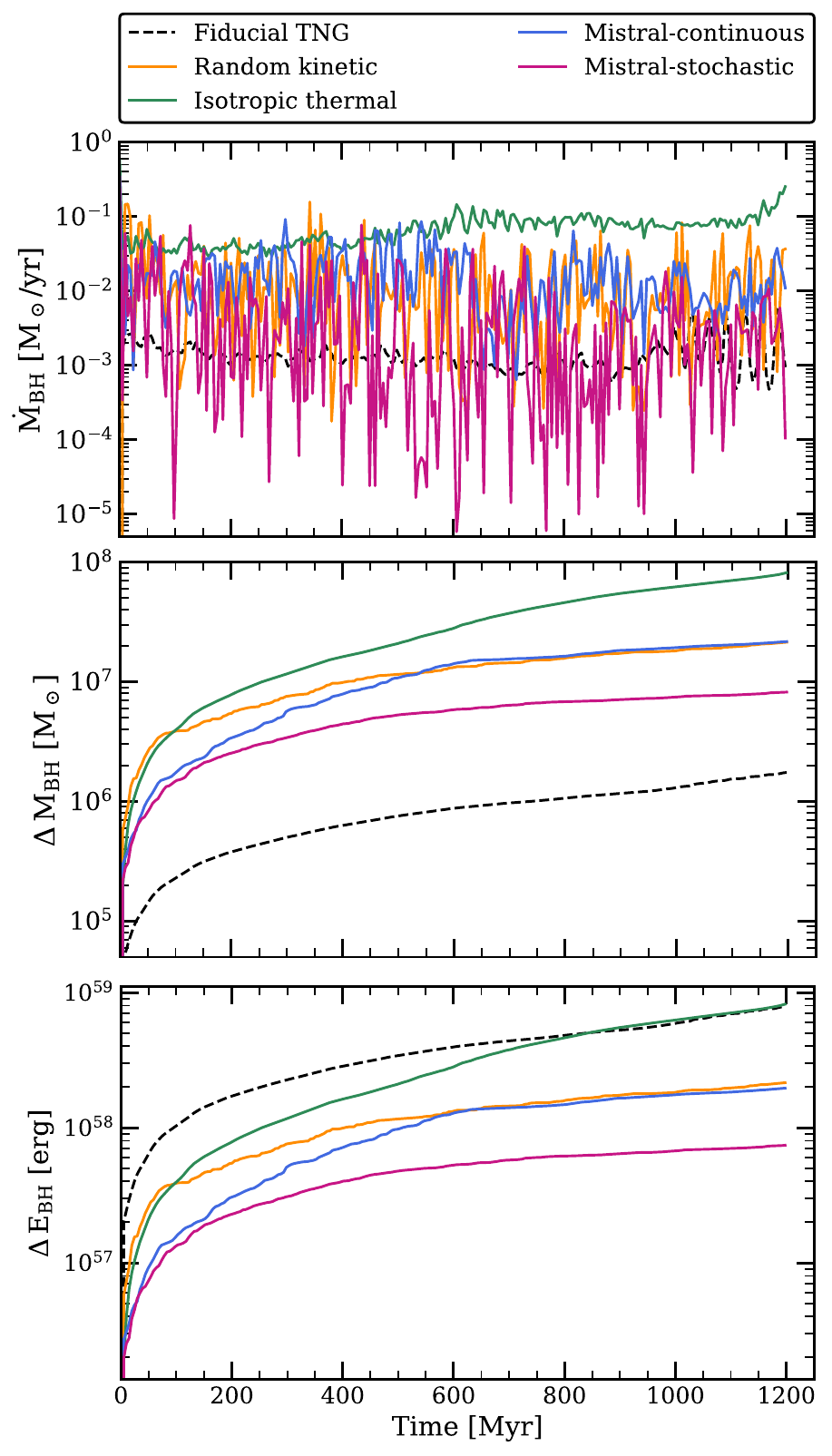}
    \caption{From top to bottom: BH accretion rate, cumulative BH mass formed and cumulative BH energy released, as a function of time. We show simulations with the \tng{} model in dashed black, and with \randomw{}, \isoth{}, \mistralC{} and \mistralS{} shown by orange, green, blue and purple solid lines, respectively.}
    \label{fig:m12bh_props}
\end{figure}

Focusing on stellar properties first, we find that \mistralS{} suppresses star formation more significantly than other models, while \mistralC{}, in contrast, globally leads to the highest SFR after the first 200 Myr. This can be directly linked to the gaseous content in the disk, which has a time evolution similar to that of the SFR. Despite having the highest stellar mass, the simulation with \mistralC{} also has the highest gas fraction of the six runs. With this feedback model, the variation of $f_{\rm gas,disk}$ with time reflects the fact that the dense outflowing gas accelerated by AGN feedback rapidly falls back to the galaxy, episodically providing more fuel for star formation. By construction, \mistralC{} injects momentum radially away, which creates density contrasts in the ISM and leads to gas compression that triggers star formation compared to the other models. In particular, the galaxy with \mistralC{} has higher SFRs and gas fractions than the simulation without BH physics. We note that our results would likely differ if using different star formation and resolved multi-phase ISM prescriptions (which we will investigate in a forthcoming study). With this setup, \mistralC{} acts as positive feedback, leading to a stellar mass formed 1.2 times higher than in the absence of AGN feedback. Given our parametrization, the simulation with the \isoth{} model produces very similar SFR and gas fraction as the \nobh{} simulation. With its isotropic release of thermal energy that gently heats up the ISM gas, the \isoth{} simulation produces lower disk gas fractions than \mistralC{}, but forms almost as many stars as the \nobh{} simulation. These quantities are reduced in the \tng{} simulation, as a result of the higher BH energy released (see also Fig.~\ref{fig:m12bh_props}), reducing the stellar mass formed by a factor of 1.6 compared to the simulation without BH physics. Because a similar star formation suppression is achieved with the \randomw{} model with less BH energy released, this shows that a larger amount of thermal energy is needed to achieve the same effect on star formation as kinetic winds (but the impact on gas properties and BH growth differs with the BH energy deposition scheme). Finally, being more efficient at ejecting gas out of the disk, \mistralS{} has even lower $f_{\rm gas,disk}$ and SFR, forming 1.8 times less stellar mass than the counterpart \nobh{} run. We note however that these differences in stellar mass formed remain negligible compared to the total stellar mass of the galaxy, which varies by no more than 2\% during the 1200 $\rm Myr$ of the simulations.

AGN feedback does not only impact star formation but also BH accretion. Fig.~\ref{fig:m12bh_props} shows the black hole accretion rate, cumulative black hole mass accreted, and cumulative BH energy released. Overall, the \randomw{} and the two \mistral{} models produce a more bursty BH accretion history than the \isoth{} and \tng{} models. Given our choice of efficiency parameters (see Sec.~\ref{subsec:gal-ics}), we remind the reader that the fraction of inflowing gas mass accreted by the BH corresponds to 90 percent in the \isoth{} and \randomw{} simulations, to 50 percent with \mistral{} and, by default, to 80 percent with the \tng{} model. This explains why the BH accretion rates are consistently higher with the \isoth{} model than in the other simulations, despite releasing more BH energy than \mistral{}, for instance. Unlike with the deposition of kinetic energy, the thermal energy released by the BH with the \isoth{} model is more prone to being radiated away in dense gas, which reduces the actual feedback efficiency \citep{Weinberger2018}. Therefore, with the low feedback coupling efficiency adopted in our simulation, the \isoth{} model is inefficient at regulating both star formation and BH growth. When this feedback model is stronger, with a higher feedback coupling efficiency in the \tng{} simulation, the higher amount of BH energy initially released efficiently regulates BH growth \citep{Terrazas2020}, and the total BH mass accreted is almost 50 times lower. Just as it is more efficient at regulating star formation, the \randomw{} model leads to lower BH accretion rates and releases less energy than the \isoth{} model, despite having the same feedback coupling efficiency.

Similarly as for the SFR, the simulation with \mistralC{} has higher BH accretion rates than the one with \mistralS{}, as the former retains more gas near the BH. This is also because \mistralS{} is more efficient at expelling and heating the gas surrounding the BH, leading to lower BH accretion rates. Compared to the simulation with the \isoth{} model, the integrated BH mass accreted differs by a factor of 3.8 and 10 for \mistralC{} and \mistralS{}, respectively, such that the BH eventually reaches a mass 1.8 lower with \mistralC{}, and 2.3 times lower with \mistralS{}. By the end of the simulation, the BH reaches a mass of $5.1\times10^7$, $7.1\times10^7$, $1.3\times10^8$, $7.1\times10^7$ and $5.8\times10^7\rm\,M_\odot$ with the \tng{}, \randomw{}, \isoth{}, \mistralC{} and \mistralS{} models, respectively. Interestingly, despite releasing less BH energy over the duration of the simulation, \mistralS{} efficiently regulates star formation, resulting in the lowest overall SFR among all the models we investigated.

\subsubsection{Properties of the outflowing gas}
\label{subsec:m12outf}

The main goal of \mistral{} is to capture the consequences of the winds that are accelerated in the vicinity of the SMBH, as a result of the energy released from the accretion disk. To quantify the efficiency of AGN feedback in generating galactic winds, this section now focuses on the properties of the outflowing gas. We define gas as outflowing when it is flowing away from the galaxy center, in a radial direction (i.e. gas cells with a positive radial velocity). To compute the mass and kinetic energy outflow rates $\dot{M}_{\rm outf}$ and $\dot{E}_{\rm outf}$, we define spherical shells of width $\Delta R_{\rm shell}=2\,\rm kpc$, located at a distance $R_{\rm shell}$ between 4 and 100 kpc from the galaxy center. $\dot{M}_{\rm outf}$ and $\dot{E}_{\rm outf}$ are computed as follows, by summing the contribution of all the outflowing cells $i$, of gas mass $m_i$ and radial velocity $v_{{\rm r},i}>0$, that are intersected by the shells:

\begin{align}
    \label{eq:outf}
    & \dot{M}_{\rm outf} = \sum_i \frac{m_i v_{{\rm r},i}}{\Delta R_{\rm shell}}\,\\
    & \dot{E}_{\rm outf} = \sum_i \frac{1}{2}\frac{m_i v_{{\rm r},i}}{\Delta R_{\rm shell}} v_{{\rm r},i}^2
\end{align}

\begin{figure}
    \centering
    \includegraphics[width=\columnwidth]{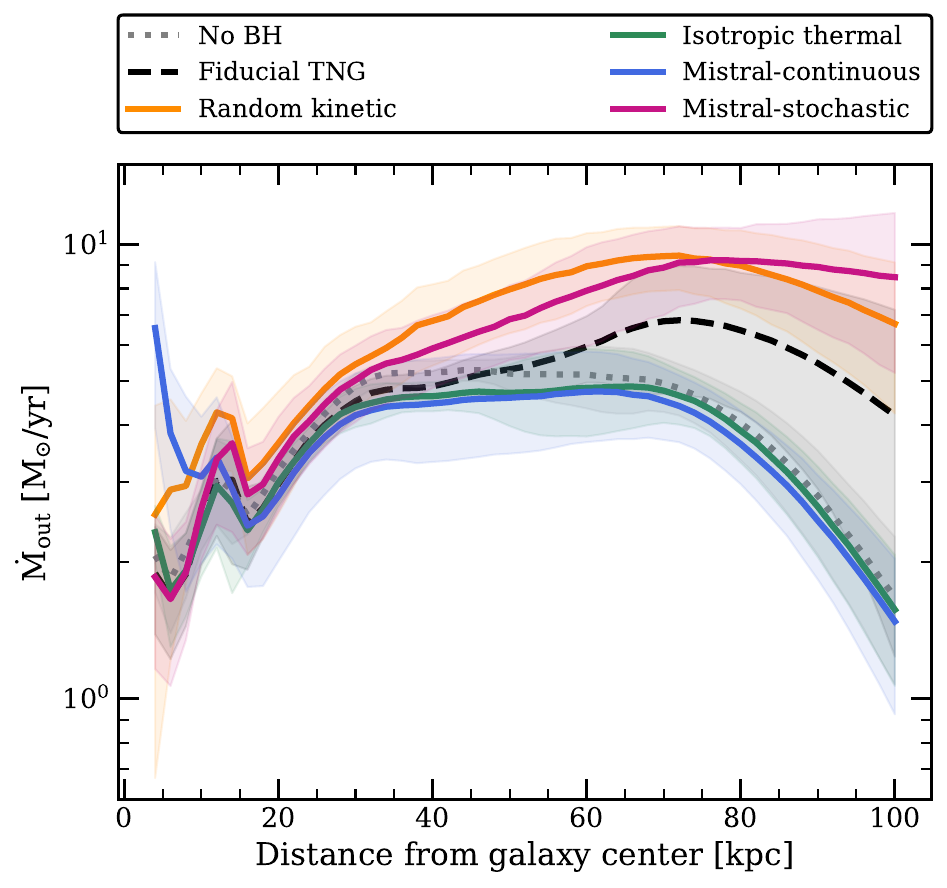}
    \caption{Mass outflow rate as a function of distance from the galaxy center, with data stacked from 200 to 1000 Myr. Colored lines represent average quantities, and the limits of the standard deviation from the mean are shown by the shaded area. Dotted grey, dashed black, solid orange, green, blue and purple lines show simulations with \nobh{}, the \tng{}, \randomw{}, \isoth{}, \mistralC{} and \mistralS{} feedback models.}
    \label{fig:m12_outrad}
\end{figure}

\begin{figure}
    \centering
    \includegraphics[width=\columnwidth]{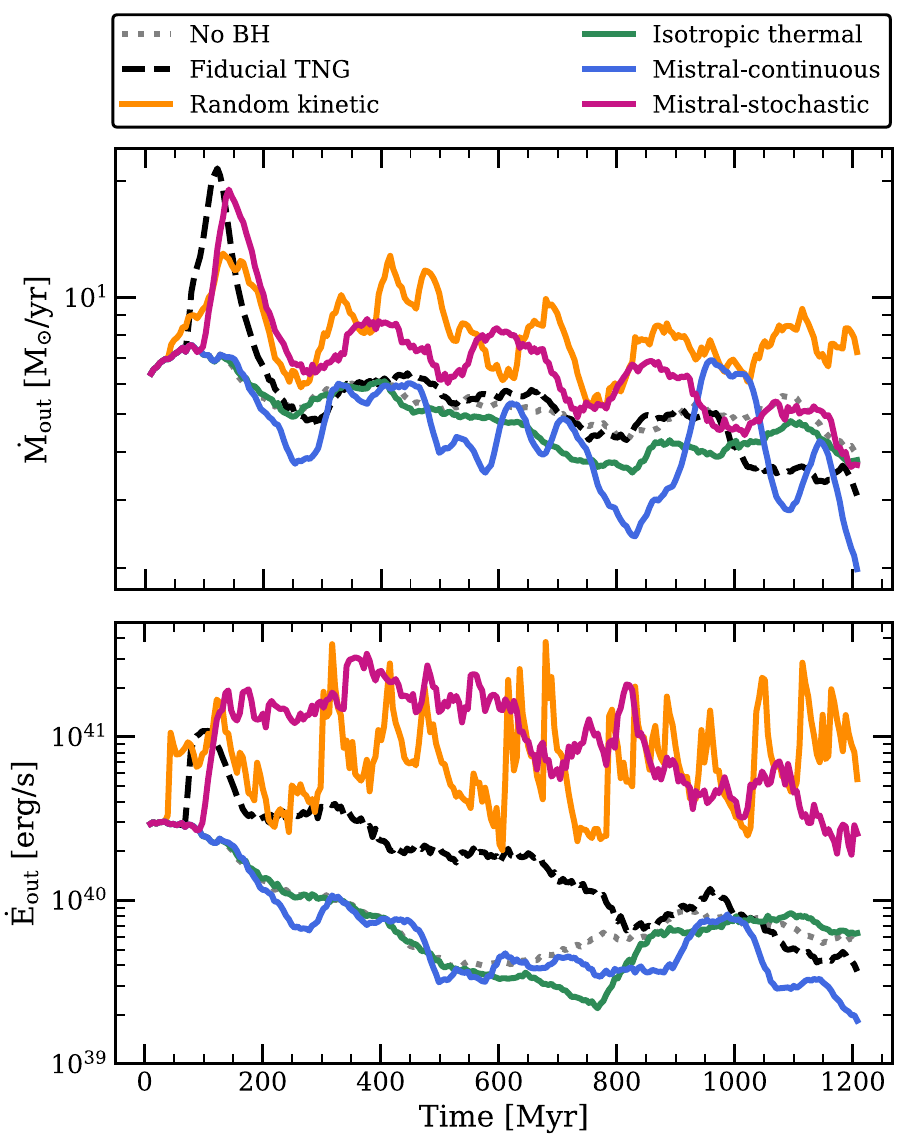}
    \caption{Mass outflow rate and kinetic energy outflow rate as a function of time, measured in a shell located at 50 kpc from the galaxy center. Dotted grey, dashed black, solid orange, green, blue and purple lines show simulations with \nobh{}, the \tng{}, \randomw{}, \isoth{}, \mistralC{} and \mistralS{} feedback models.}
    \label{fig:m12_outtime}
\end{figure}

Fig.~\ref{fig:m12_outrad} first focuses on the mass outflow rate, shown as a function of distance from the galaxy center for our six simulations. In order to reduce the noise due to the burstiness of supernova and AGN feedback, we stack 161 outputs between 200 and 1000 Myr. We show the average mass outflow rate (colored lines) together with the standard deviation (shaded areas). Because the outflow properties are calculated in spherical shells, distances below $20\,\rm kpc$ include gas that resides in the ISM. There, \mistralC{} and the \randomw{} model produce the highest mass outflow rates, by accelerating large quantities of dense gas. This is also visible in the bottom right part of the corresponding phase diagram (Fig.~\ref{fig:phase_diag}), where we see more cold, high velocity gas for these two models. At larger distances and up to $50\,\rm kpc$, \mistralC{}, the \isoth{} and the \tng{} model all produce mass outflow rates similar to the case where no BH physics is included. The outflow rates with the \isoth{} model have a smaller scatter than with \mistralC{}, due to BH accretion being less bursty. Interestingly, the mass outflow rates with the \tng{} model start to diverge at $50\,\rm kpc$ and beyond, where enough BH energy remains to support outflowing gas (see also Fig.~\ref{fig:m12_outtime}), and where the gas density is sufficiently low to avoid significant radiative losses. Beyond $20\,\rm kpc$, kinetic winds with \mistralS{} and the \randomw{} produce the highest outflow rates, reaching values close to $10\,\rm M_\odot\,yr^{-1}$ 100 kpc away from the galaxy, compared to $1-2\,\rm M_\odot\,yr^{-1}$ for \mistralC{} and the \isoth{} model. While the mass outflow rates start to drop when $R_{\rm shell} \geq 70-80\,\rm kpc$ with all the other models, this is not the case with \mistralS{}. This is a consequence of the episodic acceleration of gas along the same direction, maintaining the highest gas outflow rates at radii between 80 and 100 $\rm kpc$ from the galaxy center. At $R_{\rm shell} = 100\,\rm kpc$, \mistralS{} therefore leads to mass outflow rates almost one order of magnitude higher than the \isoth{} model and \mistralC{}.

To better quantify the efficiency of different AGN feedback models in driving large-scale winds, Fig.~\ref{fig:m12_outtime} shows the mass outflow rate and the kinetic energy outflow rate as a function of time\footnote{We refrain from showing energy outflow rates normalised by the instantaneous AGN luminosity given the high variability of the BH accretion rates with \mistral{}. This also allows better visualization of the BH energy injection events, that happen occasionally with \mistralS{} and the \randomw{} model, unlike their continuous BH accretion.}. Both quantities are measured at $R_{\rm shell}=50\,\rm kpc$, a distance that probes the CGM, and at which the mass outflow rates with \mistralC{} and the \isoth{} model reach a plateau (as visible in Fig.~\ref{fig:m12_outrad}). 

Overall, the mass outflow rates at that distance are the highest with \mistralS{} and the \randomw{} model. Conversely, \mistralC{} and the \isoth{} model produce the lowest mass and kinetic energy outflow rates, despite the total amount of BH energy released being up to five times lower with \mistralC{}. Compared to the simulation with the \isoth{} model, the one with the \tng{} physics globally has higher mass outflow rates. Because of the large amount of BH energy released during the first $200\rm\,Myr$, we also note the presence of an initial peak in mass and energy outflow rates in this simulation. We report a similar feature with \mistralS{} and the \randomw{} model after the first AGN feedback event, despite more than 10 times less BH energy being released with \mistralS{}.

Unlike the \isoth{} and \tng{} models, the two versions of \mistral{} both lead to outflowing properties that fluctuate episodically. This can be interpreted as follows. Depending on BH accretion, \mistralC{} continuously injects momentum radially away, and this occasionally launches small-scales winds, as visible in Fig.~\ref{fig:MWmapsall}. The time needed for these galactic fountains to fall back to the galaxy leads to $\sim 100-200\,\rm Myr$ fluctuations in outflow properties, as gas has to reach the ISM and be accreted by the SMBH before AGN feedback has a chance to expel gas again. Fluctuations with a similar time delay can also be observed in galaxy properties, such as SFR, BH accretion rate and gas fraction (Fig.~\ref{fig:m12sf_props} and \ref{fig:m12bh_props}). This illustrates that there is a cycle of gas, which affects BH accretion, and consequently the injection of AGN feedback. With the \randomw{} model and \mistralS{}, the episodic launching of winds directly correlates with the stochastic injection of kinetic AGN feedback, by construction. The first AGN feedback episodes are sufficiently efficient that they remove gas from the ISM, which can also be seen in Fig.~\ref{fig:m12sf_props} and the subsequent reduction of star formation and BH growth. The following AGN feedback events then contribute to maintaining this trend. Despite the total BH energy released being the lowest with \mistralS{}, this model leads to the  highest mass and energy outflow rates (except for the last $300\,\rm Myr$ dominated by the \randomw{} model at this distance from the galaxy), which is the signature of the powerful AGN-driven winds generated by this stochastic, kinetic AGN feedback model.

To conclude, in these idealised simulations, the \randomw{} model and \mistralS{} happen to produce similar mass and energy outflow rates, and lead to a similar star formation suppression. However, the AGN winds they produce have different morphologies, which differently impacts BH growth and gas properties, both in the galaxy disc and in the CGM. Because idealised simulations exclude the complexity of the CGM and cosmic inflows, the differences we see between AGN feedback models in these idealised galaxies are amplified in a more realistic cosmic environment, which we examine in the next section.

\section{Zoom simulations of massive galaxies at cosmic noon}
\label{section:resultsz2}

In this section we show how the two versions of our model affect the evolution of massive galaxies in a more realistic cosmological context. To take into account the impact of galaxy mergers and of gas inflows from the cosmic web, we turn to cosmological simulations. For this purpose, we conduct a suite of cosmological zoom-in simulations of massive $z=2$ galaxies from the TNG100 simulation \citep{Springel2018,Pillepich2018b,Naiman2018,Nelson2018,Marinacci2018}, whose galaxy formation model has been already detailed in Section~\ref{subsec:physics-tng}. This simulation corresponds to the intermediate cosmological volume of the IllustrisTNG project, with a box size of $\rm110\,cMpc$ in width. TNG100 has a dark matter mass resolution of $\rm 7.5\times 10^6\,M_\odot$, a baryonic mass resolution of $\rm 1.6\times 10^6\,M_\odot$, an adaptive gravitational softening length of 185 comoving pc at minimum for gas and a gravitational softening length of 740 pc at $z=0$ for stars and dark matter. It adopts a \citet{Planck2016} cosmology with $\Omega_{\Lambda,0} = 0.6911$, $\Omega_{m,0} = 0.3089$, $\Omega_{b,0} = 0.0486$, $\sigma_{8} = 0.8159$, $n_s = 0.9667$ and $h = 0.6774$. Unless stated otherwise, we use the same galaxy formation models and parameters as in the TNG100 simulation. Below, we expand on the galaxies targeted for our sample of zoom simulations.

\subsection{Initial conditions and sample of zoom simulations}
\label{subsec:z2-ics}

We randomly select 12 galaxies among the most massive from TNG100 at $z=2$ (three of them are also simulated at higher resolution, see Appendix~\ref{app:res}). We additionally target three massive galaxies from TNG100 at $z=0$, that are used to calibrate $\epsilon_{\rm w}$, in order to produce realistic $z=0$ (and $z=2$) galaxy and BH properties (see Appendix~\ref{app:calib}). In the TNG simulations (and in our zoom simulations), halos and galaxies are respectively defined as friend-of-friend groups and gravitationally bound objects, and are determined on-the-fly using the \subfind{} algorithm \citep{Springel2001}. Each galaxy can then be tracked back over time via its merger tree, built in post-processing using the \sublink{} algorithm \citep{Rodriguez-Gomez2015,Nelson2015}. As summarized in Table~\ref{tab:z2sample}, at $z=2$, our 15 target galaxies have stellar masses $M_\mathrm{stellar} \geq 6\times 10^{9} \,\mathrm{M}_\odot$ and are hosted in halos with masses between $M_{\rm halo}=10^{12}-3\times10^{13}\,M_\odot$\footnote{We note that, upon re-simulation with the zoom-in technique, and even if using the same galaxy formation physics and resolution, some properties may vary from the parent galaxy to its zoom counterpart, and BH mass, stellar mass and SFR will obviously differ when using one AGN feedback model or another, as shown by the different figures from this section.}. Throughout this section, if not mentioned otherwise, stellar properties are defined from all the gravitationally bound star particles within twice the stellar half-mass radius of the galaxies. We re-simulate this sample of halos using the zoom-in technique described below, and will refer to these re-simulations as "zoom" simulations even if we keep the same resolution as TNG100.

\begin{table}
	\centering
	\caption{Properties at $z=2$ of the 15 galaxies targeted for re-simulation, as obtained from the snapshot 33 ($z=2$) of the TNG100 simulation. From left to right: GrNr: halo group number of the galaxy, $M_{\rm halo}$: total mass of the halo hosting the galaxy, $M_{\rm stellar}$: stellar mass within twice the stellar half-mass radius, $\rm sSFR$: specific star formation within twice the stellar half-mass radius, $M_{\rm BH}$: BH mass. Starred halo numbers correspond to objects that are evolved down to $z=0$.}
	\label{tab:z2sample}
	\begin{tabular}{ccccc}
		\hline
		GrNr & $M_{\rm halo}$ & $M_{\rm stellar}$ & $\rm sSFR$ & $M_{\rm BH}$\\
	    at $z=2$ & [$\rm 10^{12}\,M_{\odot}$] & [$\rm 10^{10}\,M_{\odot}$] & [$\rm yr^{-1}$] & [$\rm 10^{8}\,M_{\odot}$]\\
		\hline
		$631^{*}$ & $1.1$ & $0.6$  & $1.6\times10^{-9}$  & $0.03$ \\
        511 & $2.3$ & $4.8$  & $3.5\times10^{-10}$ & $0.9$ \\
        449  & $1.6$ & $4.2$ & $3.5\times10^{-12}$ & $4.1$ \\
        306  & $2.7$ & $5.5$ & $5\times10^{-10}$ & $2.5$ \\
        200  & $3.2$ & $5.5$ & $2\times10^{-11}$ & $5.1$ \\
        190  & $4.3$ & $12.7$ & $0$ & $5.7$ \\
        $123^{*}$ & $3.5$ & $2.3$  & $1.2\times10^{-9}$  & $1.8$ \\
        113  & $6.6$ & $7.96$ & $1.3\times10^{-13}$ & $8.9$ \\
        61  & $8.8$ & $12.1$ & $3.3\times10^{-14}$ & $6.2$ \\
        $57^{*}$  & $8.4$ & $11.2$ & $10^{-10}$          & $6.3$ \\
        55  & $6.1$ & $12.8$ & $9.4\times10^{-15}$ & $7.5$ \\
        30  & $10.0$ & $15.8$ & $3\times10^{-11}$ & $11.4$ \\
        18  & $17.6$ & $32.9$ & $4.7\times10^{-11}$ & $15$ \\
        10  & $26.3$ & $20.2$ & $1.5\times10^{-12}$ & $8.7$ \\
        6  & $31.5$ & $30.1$ & $1.7\times10^{-12}$ & $18.2$ \\
		\hline
	\end{tabular}
\end{table}

For each selected halo, we first extract the list of dark matter particles at snapshot 33 from TNG100, corresponding to our target final redshift $z=2$. This is with the exception of halos with group number GrNr 631, 123 and 57, for which ICs are extracted at $z=0$ (snapshot 99), their final redshift (see Appendix~\ref{app:calib}). The spatial distribution of these particles is then located back in time, down to the initial snapshot zero (which corresponds to $z=20$), in order to identify the Lagrangian region from which the halo formed. We finally apply to this particle set a perturbation field following the \citet{Zeldo1970} approximation, in the same way as done in \ngenic{} \citep{Springel2015}, using the same cosmological parameters and power spectrum as used in the TNG100 simulation. The initial conditions generated consist of a list of dark matter particles (from which gas cells are internally generated by \arepo{}), at TNG100 resolution within the region centered on the target halo, and at a lower resolution outside, such that the computational load is exclusively focused on the region of interest. Within this zoom region, the highest resolution DM particles have a mass of $7.5\times10^6\,\rm M_\odot$, and the DM resolution progressively coarsens away from the region of interest down to $2\times10^{12}\,\rm M_\odot$. The central, "zoom" region of interest is defined as the sphere containing 4 times more DM particles than that belonging to the virial radius $R_{\rm vir}$ of the target halo at the final redshift. This extends the resolution around the halo's immediate surrounding (up to $\simeq 1.6 \times R_{\rm vir}$) in order to minimize any contamination by low resolution DM particles. 

\subsection{Impact of AGN-driven winds on z=2 massive galaxies}
\label{subsec:z2-results}

In this section, we employ our suite of cosmological simulations to investigate the effects of AGN winds on massive galaxies. We focus on BH and galaxy properties at $z=2$, which is the most promising redshift for probing the impact of AGN winds driven by radiatively efficient accretion. Indeed, this redshift, also referred to as "cosmic noon", corresponds to the epoch when both star formation and SMBH growth are at their peaks \citep{Madau&Dickinson2014}. Therefore, AGN feedback is expected to be particularly important, by generating winds that can impact star formation and gas fraction. We compare the two versions of \mistral{} to the \quasar{} model, in order to isolate the impact of AGN feedback in the radiatively efficient regime of SMBHs. Unlike the \isoth{} model used in Section~\ref{section:resultsm12}, the \quasar{} model includes the same fiducial physical models and parameters as in the TNG simulations, and we only disable the switch for the TNG \randomw{} model at low Eddington ratios. We also provide a comparison of \mistral{} to the \tng{} AGN feedback model where the kinetic winds are enabled at low Eddington ratios. In these two sets of simulations, we use the same radiative and coupling efficiency parameters as in TNG (unlike what was done in Section~\ref{section:resultsm12}). When using the \quasar{} model, the fraction of rest-mass accreted energy that couples to the gas is defined by the product of $\epsilon_{\rm r}=0.2$ and $\epsilon_{\rm f}=0.1$. When switching to the \randomw{} model, this fraction is variable but capped at $\epsilon_{\rm kin}=0.2$ (see Eq.9 from \citealp{Weinberger2017}). In the simulations with \mistral{}, we adopt $v_{\rm w}=10^4\,\rm km\,s^{-1}$ and $\epsilon_{\rm w}=10^{-3}$, which corresponds to 36\% (respectively 64\%) of the inflowing gas mass being accreted (kicked as a wind). With both \mistralC{} and \mistralS{}, this choice of parameters gives reasonable results in terms of stellar and black hole masses compared to $z=0$ observations (see Appendix~\ref{app:calib}).

We first analyze some of the main galaxy properties compared with observationally derived scaling relations in Section~\ref{subsec:z2-results1}. We then look at the impact of AGN feedback on gas fractions in Section~\ref{subsec:z2-results2} and \ref{subsec:z2-results3}, and on gas flows in Section~\ref{subsec:z2-results4}.

\subsubsection{Stellar mass, SFR, BH mass and BH accretion rate at z=2}
\label{subsec:z2-results1}

\begin{figure}
    \centering
    \includegraphics[width=\columnwidth]{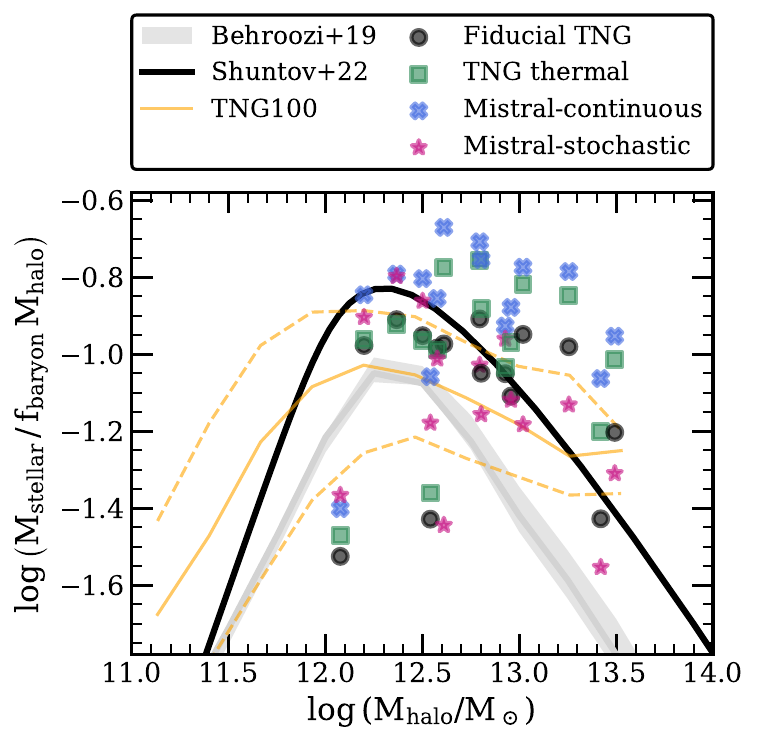}
    \caption{Stellar mass to halo mass relation at $z=2$, for zoom simulations with the \tng{} physics (black), the \quasar{} model (green), \mistralC{} (blue) and \mistralS{} (purple). We also show the stellar to halo mass relation at $z=2$ from \citet{Behroozi2019} with a grey line and shaded area, and from \citet{Shuntov2022} with a black line. To enhance clarity, the x-axis shows halo masses from zoom simulations with the \tng{} physics, such that, for each galaxy, all four models are displayed at the same x-coordinate. The solid and dashed orange lines correspond to the mean and $1 \sigma$ scatter from the TNG100 simulation, shown for reference.}
    \label{fig:smhmz2}
\end{figure}

We first focus on how AGN feedback regulates star formation. Fig.~\ref{fig:smhmz2} shows the stellar mass to halo mass (SMHM) relation, where the y-axis is defined as the ratio between the stellar mass and the theoretical baryonic mass $f_{\rm baryon}\,M_{\rm halo}=0.17\,M_{\rm halo}$. In addition, we compare our data with observational estimates at $z=2$ from \citet{Behroozi2019} and \citet{Shuntov2022}. As they do, in this section, we use the virial overdensity halo mass definition \citep{Bryan&Norman1998}. With a few exceptions, all galaxies lie around or above the expected relations\footnote{Simulations with the lowest stellar masses correspond to the galaxies hosted in halo GrNr 190 with \mistralS{} and in halo GrNr 123 with the \tng{} and the \quasar{} models.}. The two lowest mass halos (with GrNr 631 and 511) that are located before the peak of the SMHM relation, with $M_{\rm halo}<10^{12.3}\,\rm M_\odot$, have higher stellar masses with \mistral{} than with the TNG AGN feedback models. This trend changes when halo mass increases. \mistralC{} (blue) and the \quasar{} model (green) lead to the highest stellar masses, especially at the most massive end. Enabling the \randomw{} feedback model in the \tng{} zooms (black) reduces this tension, and \mistralS{} (purple stars) shows good agreement with the empirical relation from \citet{Shuntov2022}, at any halo mass. From this analysis, we conclude that the \quasar{} mode alone and \mistralC{} are both inefficient at regulating star formation in the most massive galaxies, the regime where AGN feedback is thought to have a dominant role. This motivated the inclusion of a complementary, more efficient AGN feedback mode in the TNG simulations, which is achieved with the \randomw{} model when BHs have sufficiently low Eddington ratios given their mass (following Eq.~\ref{eq:chi}, \citealp{Weinberger2017}). Here, our \mistralS{} model has a similar effect on galaxy growth, and works self-consistently across the whole halo mass range explored without the need to invoke a BH mass dependent feedback scheme.

\begin{figure}
    \centering
    \includegraphics[width=\columnwidth]{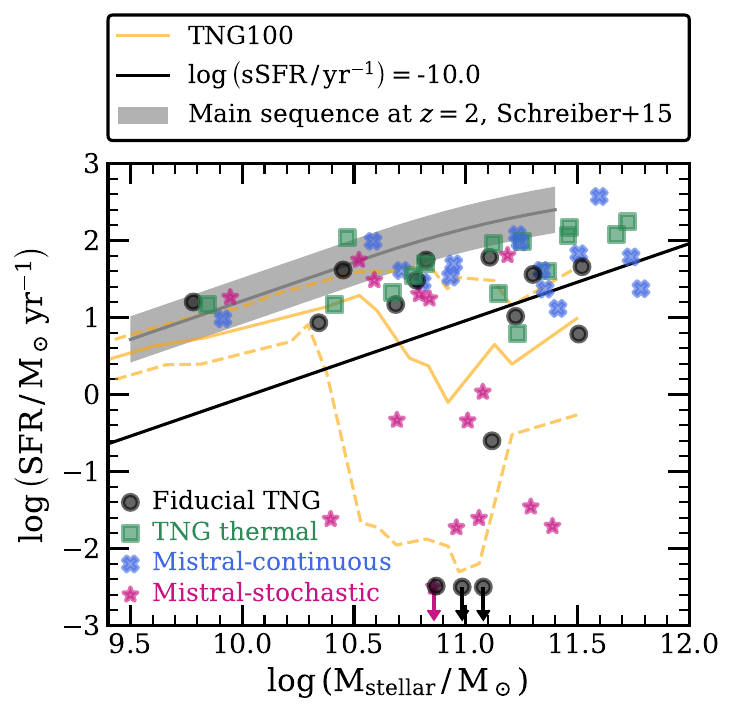}
    \caption{Star formation rate as a function of stellar mass at $z=2$ for zoom simulations with the \tng{} physics (black), the \quasar{} model (green), \mistralC{} (blue) and \mistralS{} (purple). SFR values that are lower than the limit of the plot (with \tng{} and \mistralS{}) are denoted with downwards arrows. The grey line and shaded region show the observationally derived $z=2$ main sequence and its $1 \sigma$ scatter, as taken from \citet{Schreiber2015}, and the black line corresponds to the $z=2$ sSFR quiescent threshold defined e.g. in \citet{Franx2008} as the Hubble time divided by 0.3. The solid and dashed orange lines correspond to the mean and $1 \sigma$ scatter from the TNG100 simulation, shown for reference.}
    \label{fig:sfrz2}
\end{figure}

We report a similar result in Fig.~\ref{fig:sfrz2}, which shows the SFR as a function of stellar mass. We show the star formation main sequence at $z=2$, as derived from \citet{Schreiber2015}, which is reproduced by the TNG100 simulation. The black line corresponds to the $z=2$ quiescent threshold of $\rm sSFR=10^{-10}\,yr^{-1}$, defined as the Hubble time divided by 0.3, following \citet{Franx2008}. At this redshift, almost all galaxies can be considered as star-forming when using the \quasar{} model alone, with only one galaxy having a SFR below the quiescence threshold. This number increases to two with \mistralC{}, which similarly only suppresses the SFR for some of the most massive galaxies. However, as shown in Fig.~\ref{fig:smhmz2}, this is not sufficient to produce realistic galaxy masses at $z=2$. Quenching massive galaxies is easier when enabling the \randomw{} model at low Eddington ratios in the \tng{} runs, with 6 galaxies out of 15 that drop below the quiescence limit. \mistralS{} is even more efficient at suppressing star formation, with 9/15 galaxies quenched at $z=2$. Notably, \mistralS{} efficiently acts over a wider range of galaxy mass compared to the \tng{} model, extending quenching to galaxies with lower stellar masses (down to $M_{\rm star}=10^{10.4}\,\rm M_\odot$) than any of the other AGN feedback models. While \mistralS{} yields a quenched fraction of 60 per cent at $z=2$ in our sample, this value lies within the broad observational range of 20–70 per cent reported for similar stellar masses \citep[e.g.][]{Muzzin2013,Pandya2017,Weaver2023}. In forthcoming work, we will explore quenching fractions further with a larger simulation sample.

\begin{figure}
    \centering
    \includegraphics[width=\columnwidth]{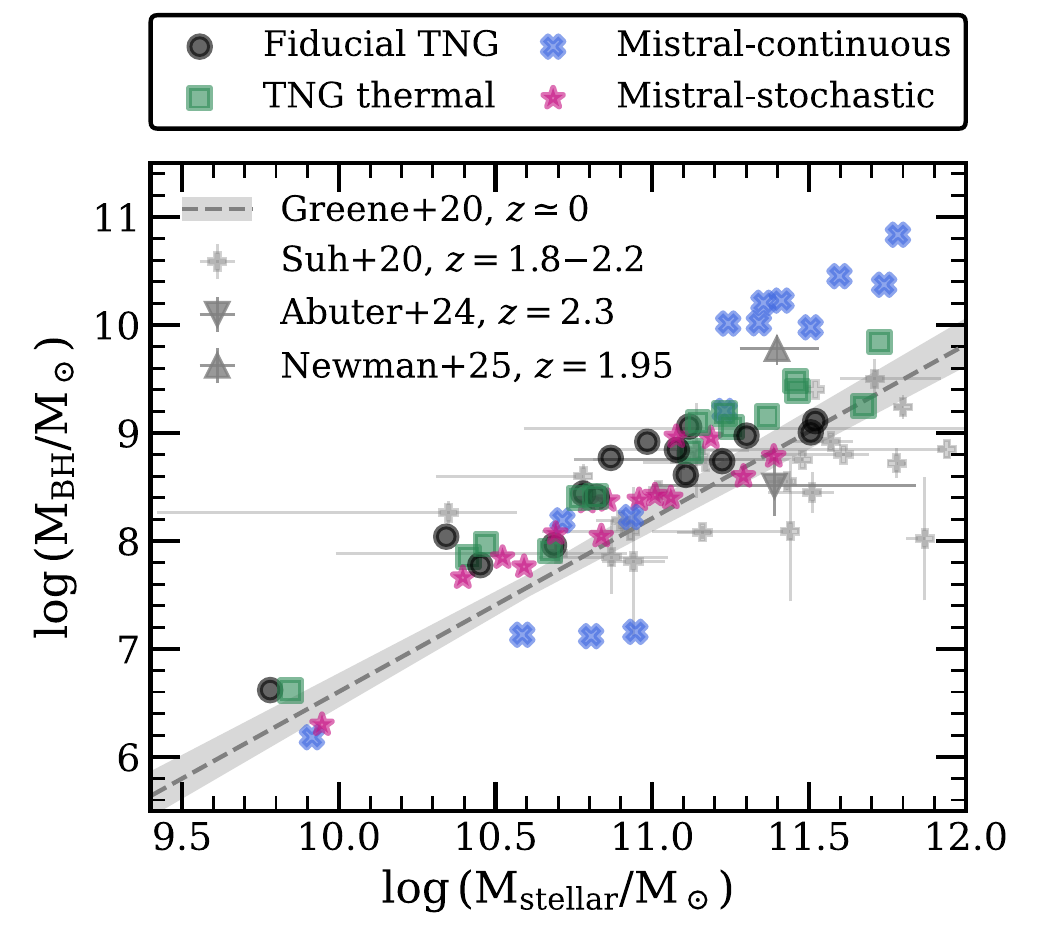}
    \caption{BH mass versus stellar mass at $z=2$, for zoom simulations with \tng{} (black), \quasar{} (green), \mistralC{} (blue) and \mistralS{} (purple) AGN feedback models. We show the local $z=0$ black hole-stellar mass relation from \citet{Greene2020}, and $z\simeq2$ observational estimates from \citet{Suh2020,Abuter2024,Newman2025}.}
    \label{fig:mbhz2}
\end{figure}

AGN feedback is expected to influence not only the growth of its host galaxy but also the evolution of its SMBH. Fig.~\ref{fig:mbhz2} shows the BH mass versus stellar mass at $z=2$, compared to two direct dynamical measurements of BH mass at $z\simeq2$ from \citet{Abuter2024} and \citet{Newman2025}. We also include observational estimates at $z=1.8$–$2.2$ from \citet{Suh2020}, though we caution that these are based on indirect methods (such as virial estimators applied to broad-line AGNs) which are subject to significant systematic uncertainties \citep[e.g.][]{Bertemes2025}, such that their use for comparison should be treated with care. Finally, we include the $z=0$ BH–stellar mass observational relation from \citet{Greene2020}, supported by observational evidence suggesting that this relation has not evolved strongly, if at all, since $z=2$ \citep[e.g.][]{Suh2020,Tanaka2025}.

Acknowledging these uncertainties, we find that with \mistralC{}, BHs are respectively under and over-massive in low and high-mass galaxies, compared to these observations. However, we note that this result depends on the value of $\epsilon_{\rm w}$, chosen as the best compromise for producing realistic stellar and black hole masses at $z=0-2$ with both \mistralC{} and \mistralS{} (Fig.~\ref{fig:mscalingz0} and \ref{fig:mscalingz2}). In Appendix~\ref{app:calib}, we show with 3 galaxies (GrNr 631, 123 and 57 at $z=2$) that a higher value of $\epsilon_{\rm w}$ tends to produce too low BH masses with both versions of \mistral{} at $z=2$, and with \mistralS{} at $z=0$. Conversely, values giving a better match with the observations at $z=2$ also produce overmassive BHs at $z=0$. With the value of $\epsilon_{\rm w}$ adopted, \mistralC{} therefore prevents BH growth in the lowest mass galaxies from our sample, and fails at regulating BH and galaxy growth in the most massive ones.

Interestingly, using the \quasar{} feedback alone leads to both higher stellar and black hole masses than with the \tng{} physics, in such a way that the BH-galaxy co-evolution remains broadly consistent with the observational estimates. BH masses with \mistralS{} are also nicely consistent with these observations, and globally leads to slightly lower BH masses than the \tng{} zoom simulations. 

\begin{figure}
    \centering
    \includegraphics[width=\columnwidth]{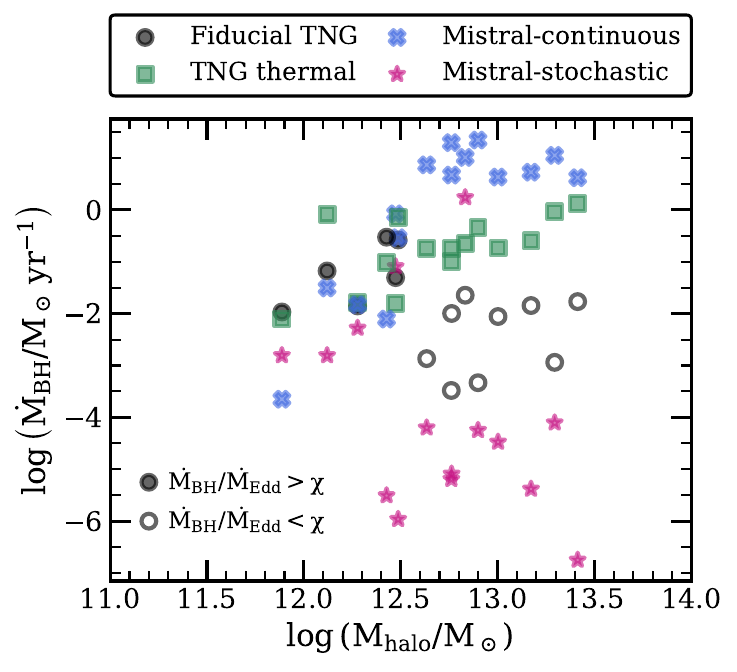}
    \caption{Black hole accretion rate as a function of halo mass at $z=2$, for zoom simulations with the \tng{} physics (black), the \quasar{} model (green), \mistralC{} (blue) and \mistralS{} (purple). With the former, empty markers indicate that the TNG \randomw{} model is active, as determined by $\dot{M}_{\rm BH}/\dot{M}_{\rm Edd}<\chi$ (see Eq.~\ref{eq:chi}). As for Fig.~\ref{fig:smhmz2}, the x-axis shows halo masses from zoom simulations with the \tng{} physics, in order to display our four models at the same x-coordinate.}
    \label{fig:bharz2}
\end{figure}

To further evaluate the impact of AGN feedback on BH growth, Fig.~\ref{fig:bharz2} shows the BH accretion rate at $z=2$, as a function of halo mass. For $M_{\rm halo}<10^{12.5}\,\rm M_\odot$, the four models lead to similar BH accretion rates, with \mistral{} producing slightly lower values. This indicates that, in this regime, both variants of our model regulate BH growth at $z=2$ at least as efficiently as, if not more efficiently than, the TNG AGN feedback prescriptions. In more massive halos, a clear trend emerges: \mistralC{} produces the highest BH accretion rates, followed by the \quasar{} model, with the \tng{} model leading to even lower values. Notably, at $z=2$, all SMBHs in halos with $M_{\rm halo}>10^{12.5}\,\rm M_\odot$ have $\dot{M}_{\rm BH}/\dot{M}_{\rm Edd}<\chi$ (where $\chi$ is defined by Eq.~\ref{eq:chi}), indicating that AGN feedback operates following the \randomw{} model. Finally, in these most massive halos, \mistralS{} generally results in the lowest BH accretion rates at this redshift, which is also true for the entire halo mass range explored (with the exception of a few objects).

To summarise, Fig.~\ref{fig:smhmz2}, \ref{fig:sfrz2}, \ref{fig:mbhz2}  and \ref{fig:bharz2} show that the \quasar{} feedback alone and \mistralC{} are generally less efficient at regulating black hole and stellar masses in massive galaxies. \mistralS{} better matches observational estimates at all halo masses, and is sufficiently strong to quench more than half of the galaxies in our sample by $z=2$. We note, however, that this study focuses on massive systems, that would be the progenitors of massive quiescent galaxies at $z=0$. In future work, we will explore how \mistral{} would regulate such galaxies as well as lower-mass objects at lower redshift, especially when the SMBH is in the radiatively inefficient regime.

\subsubsection{Cold gas fraction}
\label{subsec:z2-results2}

\begin{figure}
    \centering
    \includegraphics[width=\columnwidth]{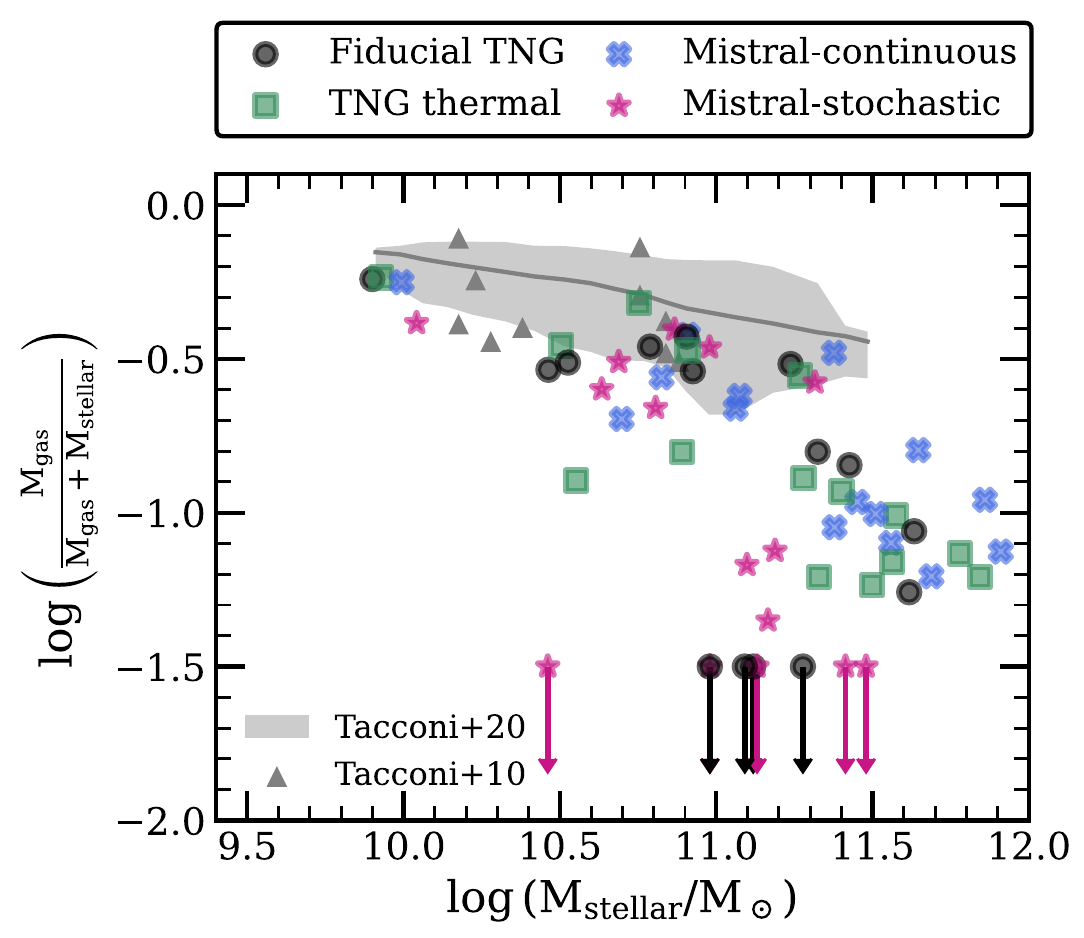}
    \caption{Cold gas fraction versus stellar mass at $z=2$, for simulations with the \tng{} physics in black, the \quasar{} model in green, \mistralC{} in blue and \mistralS{} in purple. Both the gas and the stellar masses are measured within a sphere of radius $R_{\rm vir}/10$ centered on the galaxies. Gas is defined as cold if it has a temperature $T\leq10^5\,\rm K$ and a density $\rho \geq 2\times10^{-25}\,\rm g\,cm^{-3}$. Arrows denote gas fractions below the range of the plot. The grey markers and shaded area show observations of star-forming galaxies at $z=2$ compiled by \citet{Tacconi2010} and \citet{Tacconi2020}.} 
    \label{fig:fgascold}
\end{figure}

We now want to study how AGN feedback impacts the gas content in and around galaxies. We start by illustrating how the cold gas fraction (defined as the cold gas to baryonic mass ratio) varies as a function of stellar mass in Fig.~\ref{fig:fgascold}. We measure the cold gas and stellar masses within the galaxy radius, defined as 10 per cent of the virial radius \citep[e.g.][]{Oser2010}. Gas is defined as cold if it has a temperature $T\leq10^5\,\rm K$ and a density $\rho \geq 2\times10^{-25}\,\rm g\,cm^{-3}$ (or equivalently $n_{\rm H}\geq0.13\,\rm cm^{-3}$), which corresponds to the thresholds used to define star-forming gas in the TNG simulations. For reference, we include observations of star-forming galaxies at $z=2$ from \citet{Tacconi2010} and \citet{Tacconi2020}. In observational data, cold gas is usually traced by molecular gas. This is not modelled in our simulations, that rely on a two-phase ISM subgrid prescription instead \citep{Springel&Hernquist2003}. Nevertheless, there is fairly good agreement between observations and gas fractions measured in our simulations, especially at the lower mass end, irrespective of the AGN feedback model used. Larger differences between models are seen for the most massive galaxies, for which the cold gas fraction is expected to decrease as galaxies become increasingly quenched. When $M_{\rm star}\geq10^{11}\,\rm M_\odot$, the \quasar{} model and \mistralC{} globally lead to the highest gas fractions. With the \tng{} physics, the cold gas fraction significantly drops for some of the galaxies, that correspond to the galaxies with the lowest SFR reported in Fig.~\ref{fig:sfrz2}. In these objects, the \randomw{} model of the low-accretion BH mode is enabled, which efficiently removes copious amounts of gas from the galaxies and prevents gaseous inflows. Finally, the impact of \mistralS{} on gas fraction is even more dramatic, which is also consistent with the larger number of massive quenched galaxies at $z=2$ produced with this model. The general picture depicted by Fig.~\ref{fig:fgascold} can also be related to the phase diagrams shown in Section~\ref{subsec:m12props} for our idealised galaxy simulations (Fig.~\ref{fig:phase_diag}): with the \quasar{} model and \mistralC{}, galaxies retain more cold gas in their ISM, while a significant fraction of this gas phase is heated, efficiently ejected out of the dense ISM and prevented from being re-accreted with \mistralS{}, especially in the highest-mass galaxies.

\subsubsection{Hot gas fraction}
\label{subsec:z2-results3}

AGN feedback is thought to play a role in the overall baryon content of galaxy groups and clusters \citep[e.g.][]{Eckert2021,Oppenheimer2021}, by affecting the amount and thermal state of gas at these large scales, which can also be traced by the resulting thermal X-ray luminosity of the hot gas within the DM halo \citep{McCarthy2010}. Interestingly, it has been shown that the hot gas fraction can vary substantially from one simulation to another depending on the AGN feedback model adopted \citep[e.g.][]{Choi2015,Weinberger2017}, and despite the fact that these simulations all reproduce galaxy population properties such as the massive end of the local galaxy stellar mass function \citep{Eckert2021}. Therefore, the hot gas fraction is a sensitive probe to test different implementations of AGN feedback. 

\begin{figure}
    \centering
    \includegraphics[width=\columnwidth]{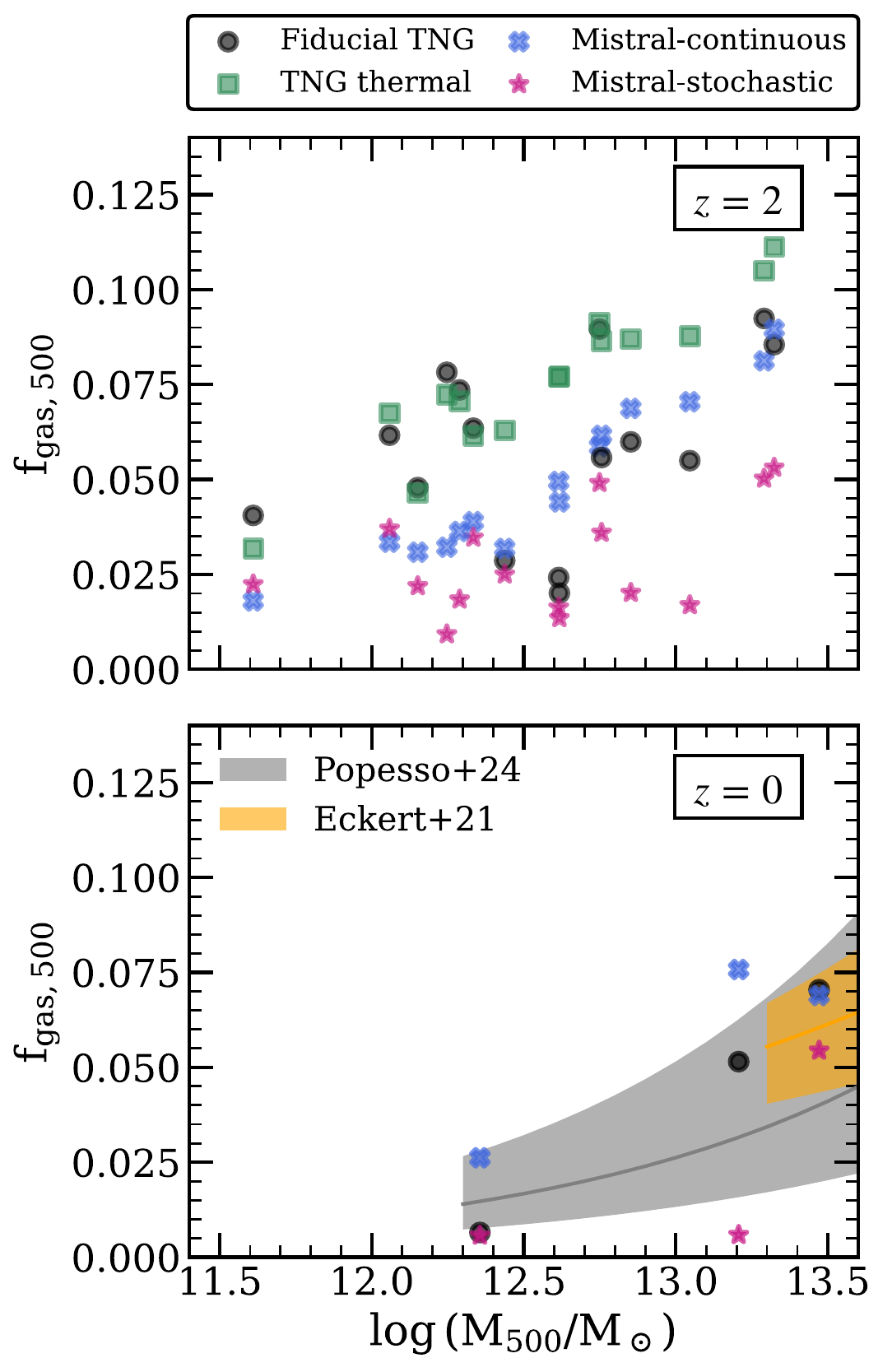}
    \caption{Hot gas fraction in the CGM at $z=2$ (top panel) and $z=0$ (bottom panel), defined as gas with $T\geq 10^{6}\,\rm K$ and $\rho \leq 2\times10^{-25}\,\rm g\,cm^{-3}$, as a function of the total mass $M_{500}$ enclosed within a sphere of radius $R_{500}$. The x-axis shows $M_{500}$ from zoom simulations with the \tng{} physics, in order to display the four setups at the same x-coordinate. The orange and grey shaded areas show the best fits to the gas fraction - halo mass relation determined from $z=0$ observations by \citet{Eckert2021} and \citet{Popesso2024}, respectively. In this work, only 3 galaxies are simulated down to $z=0$ (see Appendix~\ref{app:calib}).} 
    \label{fig:f500}
\end{figure}

For this reason, Fig.~\ref{fig:f500} shows the hot gas fraction at $z=2$ (top panel) and $z=0$ (bottom panel), as a function of the total mass $M_{500}$ enclosed within a sphere of radius $R_{500}$. The latter is defined as the region where the mean density is 500 times the critical density of the Universe, and the hot gas fraction is defined as the ratio of the hot gas mass to the total mass within $R_{500}$. We define hot gas as gas with $T\geq 10^{6}\,\rm K$ and $\rho \leq 2\times10^{-25}\,\rm g\,cm^{-3}$. For reference, we show the fit to the gas fraction - halo mass relation made by \citet{Eckert2021} from a collection of local observations, recently extended by \citet[][see their Eq. 4]{Popesso2024} to lower halo masses using eRosita observations at $z<0.2$. 

With a few exceptions, the TNG AGN feedback models and \mistralC{} tend to produce higher hot gas fractions compared to \mistralS{}. This is not only the case at $z=2$ but also at $z=0$, as shown in the bottom panel of Fig.~\ref{fig:f500}, using the three galaxies that we re-simulated down to $z=0$. With this restricted sample of objects, the \tng{} model, \mistralC{} and \mistralS{} all give results reasonably close to the fit to the observations of \citet{Popesso2024} at $z=0$. Interestingly, the \tng{} model has been shown to tend to over-predict hot gas fractions at $z=0$ in the TNG simulations \citep{Rennehan2024,Popesso2024}. Here, \mistralC{} and \mistralS{} respectively over and under-predict the hot gas fraction for one of the three galaxies. To confirm the ability of \mistral{} to reproduce realistic hot gas fractions, we would however need to repeat a similar analysis for a larger sample of massive galaxies, simulated down to $z=0$. At this redshift, where SMBHs primarily undergo radiatively inefficient accretion, we may need to recalibrate our model, as it was not designed to operate in this regime, which we will explore in future work.

Going back to $z=2$, simulations with \mistralS{} produce the lowest hot gas fractions across different halo mass ranges, compared to the other AGN feedback models tested. This shows that at $z=2$, AGN winds from radiatively efficient accretion can efficiently expel gas beyond the halo virial scale, and simultaneously regulate galaxy properties such as stellar mass, star formation rate, black hole mass and cold gas fraction. While we calibrate our model to match BH masses and stellar masses at $z=0$, \mistralS{} is not calibrated to reproduce the regulation of star formation and cold and hot gas fractions, which can be challenging for cosmological simulations \citep{Eckert2021,Rennehan2024,Popesso2024}. Thus, \mistralS{} is a promising tool for reproducing and studying the regulation of galaxy population growth due to AGN feedback.


\subsubsection{Inflows and outflows}
\label{subsec:z2-results4}

Finally, we quantify how AGN feedback regulates gas flows at different scales by focusing on mass inflow and outflow rates. To compute these two quantities, we adopt the same prescription as detailed in Section~\ref{subsec:m12outf}. Mass inflow and outflow rates are defined following Eq.~\ref{eq:outf}. They are differentiated by the sign of their radial velocity, with positive (negative) values corresponding to outflowing (inflowing) gas. To better account for the different masses and sizes of our 15 galaxies, we compute mass flow rates at the galaxy and the halo radius, respectively defined as $R_{\rm gal}=0.1\,R_{\rm vir}$ and $R_{\rm vir}$ (with $R_{\rm vir}$ varying from $56$ to $210\,\rm kpc$ at $z=2$ from the least to the most massive galaxy from our sample). For the same reason, the width of the shell within which flow rates are calculated is set to $0.01\times R_{\rm vir}$.

\begin{figure}
    \centering
    \includegraphics[width=\columnwidth]{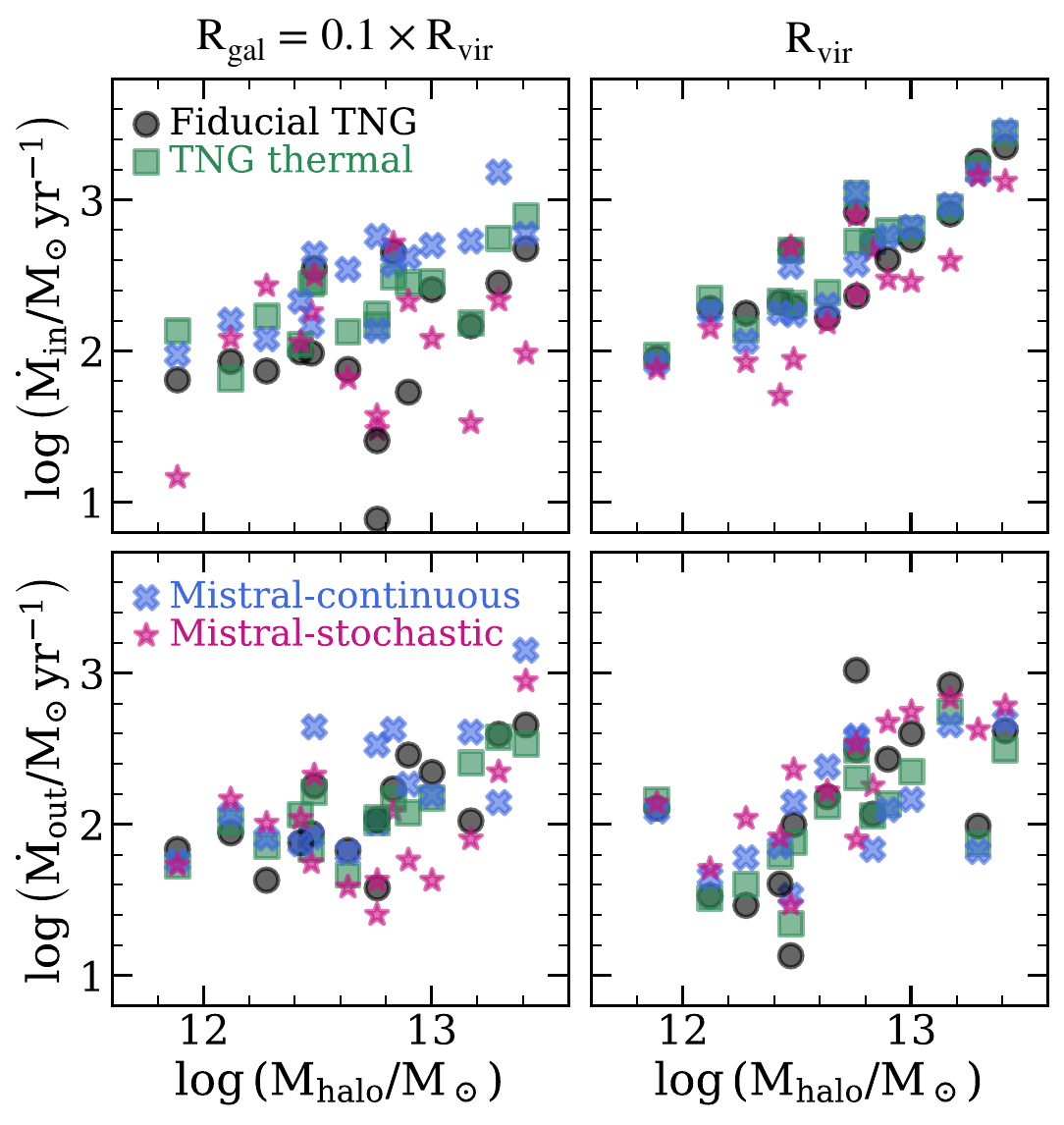}
    \caption{Mass inflow (upper plots) and outflow (bottom plots) rates at $z=2$ as a function of halo mass, measured in spherical shells of $0.01\times R_{\rm vir}$ in width, located at the galaxy radius (defined as 10 per cent of the virial radius, leftmost panels) and at the halo virial radius (rightmost panels). The x-axis shows the halo mass from zoom simulations with the \tng{} model such that inflow and outflow rates are displayed at the same x-coordinate.} 
    \label{fig:inoutz2}
\end{figure}

The resulting mass inflow and outflow rates at $z=2$ are shown as a function of halo mass in Fig.~\ref{fig:inoutz2}. At the galaxy radius, simulations with \mistralC{} have the highest mass inflow and outflow rates. This again illustrates the small-scale gaseous fountains driven with this model. With \mistralC{}, AGN feedback expels gas that rapidly falls back to the galaxy, which is notably confirmed with mass outflow rates that are lower at $R_{\rm vir}$ than at $R_{\rm gal}$. Conversely, \mistralS{} leads to the opposite behaviour for the most massive galaxies, that respectively have among the lowest mass outflow rates at $R_{\rm gal}$, but the highest mass outflow rates at $R_{\rm vir}$. This can be explained by the fact that these massive galaxies have very little gas left at $z=2$ (Fig.~\ref{fig:fgascold}), and experience fewer AGN feedback episodes as a result of the lower BH accretion rates (Fig.~\ref{fig:bharz2}). When there is enough gas left to feed the SMBH, galaxies with \mistralS{} have their outflows enhanced, which is the case for galaxies with $M_{\rm halo}\leq10^{12.6}\,\rm M_\odot$. This is also the case at any halo mass at higher redshift. For instance, we checked (but do not show) that mass outflow rates measured at $R_{\rm gal}$ are higher with \mistralS{} than with the other models at $z=3$ and $z=4$. In addition, galaxies with \mistralS{} have their inflow rates suppressed at both halo and galaxy scales. Compared to \mistralC{} and the \quasar{} model, the most massive galaxies with \mistralS{} have their inflow rates at $R_{\rm gal}$ suppressed by up to 1 dex. AGN feedback with \mistralS{} therefore prevents inflowing gas from reaching the galaxy (by providing pressure support beyond the galaxy scale), and simultaneously efficiently removes material from the halo. \mistralS{} acts simultaneously in an ejective and preventive way, just as does the model from \citet{Choi2012}, as shown by \citet{Brennan2018}, and the TNG model \citep{Zinger2020,Wright2024}. Our simulations with the \tng{} model similarly have inflows suppressed at the galaxy scale, but both the suppression of inflow rates and the enhancement of mass outflow rates at the virial radius are less pronounced than with \mistralS{}.

\section{Discussion}
\label{section:disc}

In this section, we contextualize our results by comparing them to other mechanical, kinetic AGN feedback models (Section~\ref{subsec:comparison}), before discussing the limitations of our work and potential avenues for improvement (Section~\ref{subsec:limits}).

\subsection{Comparison with AGN feedback models from the literature}
\label{subsec:comparison}

\mistral{} builds upon previous efforts to develop more physically-motivated models of AGN-driven winds. In this section, we focus on comparing \mistral{} with other AGN feedback prescriptions adopted in cosmological simulations, in order to highlight the unique features of our model.

In this study, we illustrate how the numerical AGN energy injection scheme impacts galaxy properties and evolution. In particular, we show that modelling AGN feedback via continuous thermal energy deposition, as done in the TNG quasar model, is inefficient at driving fast large-scale winds, such as those observed in some AGN \citep[e.g.][]{Bischetti2019,Bischetti2024}. This may be the consequence of artificial overcooling, which corresponds to the fact that thermal energy is quickly radiated away, if resolution is not high enough to let the energy propagate away from its injection site \citep{Weinberger2018}. To alleviate this issue and make the quasar feedback more efficient, one possibility is to increase the amount of energy injected, by introducing a duty cycle to store the AGN energy \citep{Booth&Schaye2009,Henden2018}. This is what is used in the simulations presented by \citet{Schaye2015}, who additionally switch from a continuous to a stochastic energy deposition scheme. Another solution, adopted in \mistralC{}, is to rely on (continuous) momentum injection. While this improves the situation in generating small-scale galactic winds, the model, as it is, does not reproduce realistic BH and galaxy properties at $z=2$. By combining a stochastic AGN feedback scheme with momentum injection, \mistralS{} produces large scale winds driven by BH accretion, while effectively regulating BH and galaxy evolution. Using idealised simulations of galaxy groups and clusters, \citet{Husko2024} also found that directional kinetic feedback is more efficient at removing cold gas and suppressing star formation than isotropic thermal or kinetic feedback, which is similar to what we find with \mistralS{} and the TNG \randomw{} model, when compared with the TNG \isoth{} model and \mistralC{}.

Although the TNG \randomw{} model produces similar outflow properties as \mistralS{} (Sec.~\ref{section:resultsm12}), the two models lead to quantitatively different CGM gas morphology, and \mistralS{} needs less cumulative BH energy to produce a noticeable effect on gas fraction and star formation. These differences arise from the timing and location of AGN energy deposition. Because the two models require different conditions for releasing energy, they naturally lead to different AGN feedback duty cycles. In addition, differences in the directionality of the injection affect whether energy is preferentially coupled to low-density CGM gas or to denser gas in the galaxy disk, which can occasionally occur with the TNG \randomw{} model. Together, these two aspects impact the efficiency of AGN feedback on large scales. In addition, the TNG \randomw{} model has only been used in the radiatively inefficient regime in a cosmological context, and it would likely require re-calibration in order to operate in both SMBH regimes, as \mistralS{} does. We also note that, in this study, we do not intend to quantitatively compare the AGN-driven winds generated by \mistral{} with observations of BAL outflows. We also do not assess how realistically \mistral{} operates in the radiatively inefficient regime, or in regulating galaxies within halos of $M_{\rm halo}\leq7\times10^{12}\rm\,M_\odot$, using the same parameterizations for radiatively efficient AGN feedback at different redshifts. These are all aspects that we plan to explore in detail in subsequent works.

\mistralS{} is the \arepo{} analogue of the AGN feedback model developed by \citet{Ostriker2010} and implemented by \citet{Choi2012} in \gadget{}. First, we should remind the reader that the two astrophysical codes rely on a different computational method for solving the hydrodynamical equations. \gadget{} \citep{Springel2005gadget} employs the Lagrangian SPH (smoothed-particle hydrodynamics) technique \citep{Monaghan1992}, which considers dark matter, gas and stars as particles. \arepo{} \citep{Springel2010} is a descendant of \gadget{}, which discretizes the MHD equations on a moving Voronoi mesh, combining the advantages of grid-based and SPH schemes. In both \mistralS{} and its \gadget{} equivalent, gas elements (cells with \arepo{} and particles with \gadget{}) are stochastically kicked at a pre-defined wind velocity. In addition, the model from \citet{Choi2012} shares the momentum imparted to the wind particle with up to two neighboring gas elements. All particles have the same velocity increment, corresponding to a third of the initial wind velocity. Because this correspondingly decreases the total kinetic energy released, the residual energy is then deposited in the thermal form to ensure energy conservation. This feature was primarily developed for its computational advantages: reducing the wind velocity increment limits the reduction of the numerical timestep, and sharing momentum among several particles allows the model to better capture the AGN outflows at low resolution, when the number of gas particles is insufficient to propagate the AGN wind otherwise. Thanks to its grid-based approach, \arepo{} treats energy and momentum propagation via fluxes. This allows the \arepo{} version of the model to capture the AGN wind evolution at low resolution (by enabling cell refinement, if needed), without suffering from the technical limitations of SPH prescriptions. For this reason, we do not utilize momentum sharing in our current implementation of \mistralS{}. Unlike \citet{Choi2012}, we also do not consider the radiative feedback from the SMBH X-ray radiation, which has been shown to have a negligible effect on the stellar and gas properties of massive galaxies \citet{Choi2017}.

Overall, \mistralS{} produces similar results as the model from \citet{Choi2012}. Using an idealised galaxy disk simulation, \citet{Choi2012} illustrated how their model produces higher wind velocity and energy compared to isotropic thermal feedback models. With $z=0$ cosmological zoom simulations, \citet{Choi2015,Choi2017,Choi2018} showed that their AGN feedback model leads to a realistic regulation of black hole growth, star formation, and galaxy properties such as their sizes and X-ray luminosities. Using the same model, \citet{Brennan2018} analysed mass inflow and outflow rates at galaxy and halo scales, and demonstrated the ability of the \citet{Choi2012} feedback model to launch powerful galactic winds, while simultaneously acting via preventive feedback, similar to what we find and show in Fig.~\ref{fig:inoutz2} (see their Fig.~12 and 13). Despite using different astrophysical codes and galaxy formation physics, our implementation of the model of \citet{Choi2012} therefore produces qualitatively consistent results.

Other kinetic AGN feedback models have been used in cosmological simulations, primarily for the radiatively inefficient regime. In the Horizon-AGN simulation \citep{Dubois2014}, bipolar outflows are launched with a velocity of $10^4\,\rm km\,s^{-1}$, following the model implemented in the grid code \ramses{} by \citet{Dubois2012}. Unlike \mistral{}, this feedback mode only operates when the Eddington ratio is lower than $0.01$, and quasar feedback is modelled via thermal energy injection. A bimodal feedback scheme is also adopted in the TNG simulations \citep{Nelson2019}, in which AGN feedback is modelled with a kinetic prescription only at low Eddington ratios \citep{Weinberger2017}. In addition, they do not impose any bipolarity for the wind: instead, its direction is randomly chosen at each feedback event, leading to collimated outflows that preferentially escape along the path of least resistance \citep{Nelson2019b}. With our model, the bipolarity of the AGN wind is controlled by an angular momentum weight in \mistralC{}, and by directly injecting momentum along the direction of the angular momentum of the inner disc with \mistralS{}. 

\mistralS{} is similar in some ways to the AGN feedback model in the Simba simulations \citep{Dave2019}. Simba is the first large-scale cosmological simulation that adopted a kinetic prescription for both the radiatively efficient and inefficient regimes, using the \gizmo{} mesh-free code \citep{Hopkins2015}. As we do with \mistralS{}, gas elements are ejected in a strict bipolar way, following a certain wind probability set by the desired mass loading factor (see Eq. 9 in \citealp{Dave2019}). In contrast with our work, the wind velocity varies with BH mass or Eddington ratio, depending on the feedback mode. Lower velocity winds are launched in the high-accretion BH regime (with typical velocities of a few hundreds kilometers per second), and winds as fast as $\sim 10^4\,\rm km\,s^{-1}$ (similar to the velocities we use in \mistral{}) can only be launched in the low-accretion rate regime. Also conducted with the \gizmo{} code, the galaxy zoom simulations from the \fire{} project \citep{Hopkins2014,Hopkins2018} include a kinetic model for fast nuclear winds, without distinguishing between BH accretion regimes. Similar to the wind model of \citet{Choi2012} and to \mistral{}, their model, first introduced by \citet{Hopkins2016}, assumes that a fraction of the gas accreted onto the BH is blown out as a wind at a certain velocity. However, gas particles are launched in the plane of the BH accretion disk, creating planar outflows instead of the bipolar winds obtained with \mistralS{}. More recently, \citet{Torrey2020} updated this model to inject kinetic winds using a particle spawning technique, enabling better resolution of the wind shock. In this approach, each newly spawned wind particle is given a fixed temperature, metallicity, and radially outward velocity. Using this model, \citet{Cochrane2023, Wellons2023, Byrne2024} all show that the resulting AGN outflows regulate galaxy properties, such as their size, stellar mass and morphology.

Most AGN feedback models used in cosmological simulations to date have been calibrated to regulate massive galaxies, and reproduce a range of $z=0$ observations. At low redshift, AGN feedback is expected to be dominated by a radio-jet mode, active during episodes of low accretion rates. For this reason, a wealth of theoretical studies try to capture the complex jet-ISM interaction, and its impact on galaxy growth and properties \citep[e.g.][]{Bourne&Sijacki2017,Talbot2021,Husko2022,Borodina2025}. Conversely, less effort has been directed towards the modelling of the radiatively efficient ("quasar") mode \citep[but see e.g.][]{Costa2020,Rennehan2024}, which is much more common in gas rich high-redshift galaxies.
Given the growing number of high-redshift quasar, BAL winds, and galaxy observations that challenge our models of BH and galaxy evolution \citep[e.g.][]{Carnall2023,Wang2023,Yang2023,Bischetti2024,Matthee2024}, and the difficulty simulations seem to have in producing sufficient numbers of massive quenched galaxies at $z=3-7$ \citep[e.g.][]{Valentino2023,Weibel2025,Lagos2025,Weller2025}, improving the physical fidelity of radiatively efficient AGN feedback models is extremely timely.

\subsection{Uncertainties and limitations}
\label{subsec:limits}


We now summarize some intrinsic limitations of the \mistral{} model and of our work. To reproduce the outflows thought to be generated by accretion onto SMBH, \mistral{} assumes that a fraction of the inflowing gas mass, corresponding to $M_{\rm BH,wind}$, is ejected as an AGN wind, while the rest is effectively accreted. As described in Section~\ref{subsec:mistral1}, this statement is not strictly satisfied with \mistralC{}. Instead, the AGN energy is released (in the form of momentum) into a prescribed number of BH neighbouring gas cells, for consistency with what is done (in the form of thermal energy) in the TNG quasar mode. When the total gas mass receiving the BH energy exceeds $M_{\rm BH,wind}$, this choice over-dilutes the SMBH energy, potentially affecting the strength of AGN feedback. This does not apply to \mistralS{}: the SMBH energy is distributed over the expected outflowing gas mass, provided that $M_{\rm BH,wind}$ is higher than the gas mass resolution. Therefore, with \mistralS{}, the timing of AGN feedback episodes depends on gas resolution, via the probability of momentum injection (Eq.~\ref{eq:proba-outf}). In Appendix~\ref{app:res}, we show that stellar and black hole masses at $z=3$ are barely impacted by changing the numerical resolution with \mistralS{}. However, at higher redshift ($z\gtrsim7$), the timing of the first AGN feedback event may impact the early stages of BH and galaxy growth. 

Capturing the consequences of AGN feedback on galaxy evolution also relates to the details of BH seeding and accretion, that determine the amount of SMBH energy released. In this work, BH accretion is modelled following the Bondi-Hoyle-Littleton formalism. This is a simple model, which predicts growth rates based on the SMBH mass, assuming a spherical symmetry while remaining agnostic about the angular momentum transfer of the gas. Other models, implemented in cosmological simulations, reduce the dependency of the accretion rate on the BH mass. One such example is the torque-limited accretion \citep{Hopkins&Quataert2011, AnglesAlcazar2013,AnglesAlcazar2015,AnglesAlcazar2017}, which regulates BH growth based on the fact that gas has to lose angular momentum via gravitational instabilities in order to be driven from galaxy scales down to the BH accretion disk. Together with improved BH seeding prescriptions \citep[e.g.][]{Bhowmick2024}, more realistic treatments of BH accretion would improve the physical fidelity of AGN feedback models, and enable a better interpretation of the BH-galaxy co-evolution through cosmic time.

For this study, \mistral{} is implemented and used in the framework of the TNG simulations. Like all cosmological simulations, the TNG simulations (and therefore our work) rely on a number of simplifying assumptions regarding the galaxy formation physics. Ideally, capturing galaxy evolution would involve resolving the complex, multi-phase structure of the ISM where star formation and BH accretion take place. These processes happen at sub-parsec scales that are prohibitively expensive to reconcile with cosmological volumes of several tens or hundreds of $\rm Mpc$ in width. To encapsulate the range of physical processes happening at unresolved scales, the TNG simulations couple star formation and feedback subgrid models with an effective equation of state \citep{Springel&Hernquist2003,Vogelsberger2013}. Recently, alternatives have been developed, such as \smuggle{} \citep{Marinacci2019}, a model that resolves the multi-phase structure of the ISM implemented in \arepo{}. It additionally includes an improved prescription for stellar feedback by self-consistently treating the outflows generated during supernova explosions \citep[see also][]{Smith2018}, as opposed to the TNG decoupled stellar wind-model used in our simulations \citep{Pillepich2018}. Adopting an improved model of the ISM could enhance the robustness of our conclusions, as the structure of the ISM itself can affect the AGN wind propagation, and hence impacts the outflow properties, such as their temperature and velocity \citep{Ward2024,Sivasankaran2025}.

Another important aspect for accurately simulating the interactions between AGN winds and galaxies is to capture the scales at which SMBH energy couples with the ambient gas. As already mentioned, BH accretion and wind launching happen at sub-parsec scales, a resolution very challenging to reach in combination with the $\rm Mpc$ scales of cosmological simulations. To circumvent this limitation, one promising avenue is to make use of hyper-refinement techniques, which allow one to restrict the computational load of an increased resolution to a specific location and for a limited amount of time. This is precisely the approach adopted by \citet{AnglesAlcazar2021}. Using a galaxy zoom simulation from the \fire{} project \citep{Hopkins2014,Hopkins2018}, they used a hyper-Lagrangian refinement technique to connect galaxy scales to the sub-parsec scale of the BH accretion disk. However, this is achieved for a short duration of a few $\rm Myr$, and they do not account for AGN feedback. In the spirit of increasing resolution dynamically around the SMBH, a variety of super-Lagrangian refinement techniques have been implemented and coupled to AGN feedback prescriptions \citep[e.g.][]{Curtis&Sijacki2016,Curtis&Sijacki2016b,Koudmani2019,Bourne&Sijacki2017,Beckmann2019,Bourne2019,Costa2020}. However, because of their computational cost, none of them have been applied to large-scale cosmological simulations to date, nor to extended periods of time in cosmological zoom simulations.

\section{Conclusions}
\label{section:ccl}

Building on the success of the AGN feedback model from \citet{Ostriker2010, Choi2012}, we introduce and implement the \mistral{} subgrid model of AGN winds from radiatively efficient accretion into the \arepo{} code. \mistral{} aims to capture AGN feedback from broad absorption line winds, which are driven during BH accretion by physical processes occurring at scales typically unresolved in cosmological simulations.
To compare the impact of \mistral{} to previous models that rely on isotropic thermal energy injection, we introduce two versions of \mistral{}. \mistralC{} serves as a natural extension to the continuous, isotropic thermal energy deposition scheme used at high Eddington ratios in the TNG simulations, by continuously injecting momentum in a similar spherical fashion. In the spirit of the original model developed in \gadget{} by \citet{Choi2012}, \mistralS{} imparts momentum stochastically to gas cells, with the direction of the velocity kicks being set to create bipolar winds.
We first test both versions of our model in an idealized galaxy simulation, coupling \mistral{} with the TNG galaxy formation physics models for star formation, the ISM, stellar feedback, etc. In this controlled experiment, we compare the effects of \mistral{} to those of the TNG AGN feedback models (\isoth{} and \randomw{} models), and report the following results:

\begin{itemize}
\item \textit{\mistral{} generates galaxy-scale winds, whose properties depend on the momentum injection scheme.} When momentum is injected continuously and radially with \mistralC{}, AGN winds are dense, short-lived, and form a galactic fountain. Alternatively, when momentum is stochastically imparted to gas cells with \mistralS{}, large-scale, fast bipolar winds form, creating hot and diffuse pathways through the CGM. This result contrasts with traditional AGN feedback models that rely on continuous thermal energy injection: when using the same energy coupling efficiency as with \mistral{}, the TNG \isoth{} model does not produce a noticeable AGN wind.
\item \textit{AGN winds from radiatively efficient accretion} can significantly impact galaxy evolution. By expelling dense gas that quickly falls back to the ISM of the galaxy, \mistralC{} acts as positive feedback, enhancing the SFR through higher gas fractions than the other models. Conversely, the large scale AGN winds generated with \mistralS{} quite effectively reduce the cold gas fraction, star formation, and BH accretion.
\item \textit{Episodic kinetic AGN feedback models are more efficient at generating large-scale outflows than continuous energy or momentum deposition schemes}. With \mistralS{} and the TNG \randomw{} model, the episodic AGN feedback events produce persistent winds that propagate to  large distances from the galaxy, with mass and energy outflow rates higher than those obtained with the other models. \mistralC{} and the TNG \isoth{} model lead to similar mass and energy outflow rates over time and distance, except within the inner few $\rm kpc$, where \mistralC{} accumulates gas.
\end{itemize}

In the second part of our work, we examine \mistral{}'s ability to reproduce realistic galaxies at $z=2$. For this purpose, we conduct 4 sets of cosmological zoom simulations, targeting 15 massive galaxies from the TNG100 simulation. For each galaxy, we run simulations with both versions of \mistral{}, the \quasar{} model (i.e. the \isoth{} and radiative AGN feedback models), and the \tng{} models (which include the \randomw{} model at low Eddington ratios). Isolating the \quasar{} model allows for a more direct comparison of \mistral{} with previous prescriptions that have been used to model AGN feedback in the radiatively efficient SMBH regime. Simulations with the \tng{} physics provide a more balanced comparison between \mistral{} (which covers both regimes of BH accretion) and the well tested AGN feedback model used in TNG. We mainly focus on \mistralS{}, as it produces the more realistic and interesting results. Our conclusions are as follows:

\begin{itemize}
\item \textit{\mistralS{} efficiently regulates galaxy growth across the entire range of halo masses tested, and can lead to star formation quenching at $z=2$.} Conversely, \mistralC{} and the \quasar{} model tend to over-predict the stellar mass of the most massive galaxies. With the \tng{} model, enabling the low-accretion BH mode with the \randomw{} model is essential to suppress star formation and quench galaxies at $z=2$.
\item \textit{\mistralS{} shows a broadly consistent trend of BH-stellar mass evolution, with simulated BH masses that fall within the range of observational estimates of BH masses at $z=2$.} At $z=2$, this model produces BH masses that are slightly lower than the \tng{} model, and lie slightly above the $z=0$ observational SMBH versus stellar mass relation from \citet{Greene2020}. All AGN feedback models tested, except \mistralC{}, also produce SMBH and stellar masses lying close to this observational relation. This suggests that the efficiency of \mistralC{} may need to vary with time or galaxy mass to regulate BH growth consistently across different galaxy masses.
\item \textit{\mistralS{} realistically modulates gas fractions in massive galaxies.} Our model reduces the cold gas content in the ISM of the most massive galaxies and regulates their hot gas content. The hot gas fractions at $z=0$ are broadly consistent with observational measurements, but this result, based on three simulated halos, would require a larger statistical sample for a more robust comparison. The regulation of both cold and hot gas fractions with \mistralS{} is achieved without requiring a SMBH mass-dependent AGN feedback scheme, unlike the \tng{} model. Both \mistralC{} and the \quasar{} model fail to suppress cold gas fractions and star formation in massive galaxies, and produce high hot gas fraction at the halo scale.
\item \textit{\mistralS{} acts via both preventive and ejective feedback mechanisms.} Indeed, it suppresses inflows at the galaxy and halo scales, while driving outflows up to the virial radius of the halos. \mistralC{} primarily enhances mass outflow rates in the most massive galaxies at small scales, but does not provide sufficient pressure support to affect gas inflows. Unlike the \quasar{} model alone, the \tng{} model occasionally helps in preventing gas inflows and in generating outflows, although to a lesser extent than \mistralS{} does at large scales.
\end{itemize}

The numerical details of AGN feedback modelling can profoundly impact the properties of simulated galaxies, influencing our understanding of the physics governing the evolution of galaxies. In summary, our \mistralS{} model successfully reproduces realistic stellar masses and cold gas fractions at $z=2$, and simultaneously regulates SMBH masses and hot gas fractions. It enhances outflows, removing gas from the galaxies and the halos, and suppresses inflows, bringing some galaxies to a quiescent state by $z=2$. \mistralS{} works self-consistently across the entire halo mass range explored, capturing the consequences of AGN feedback in the radiatively efficient regime, while avoiding the BH-mass dependent TNG feedback scheme. Thanks to this success, \mistralS{} has been shown to be a promising tool for deciphering the role of AGN winds from radiatively efficient accretion in shaping galaxy properties through cosmic time. Our model paves the way for making robust predictions for the impact of AGN-winds using cosmological simulations. In a following work, we will create synthetic spectra following the approach of \citet{Hirschmann2017,Hirschmann2019,Hirschmann2023}, which will be crucial to interpret the wealth of high-redshift quasar and galaxy observations made with current and upcoming facilities such as JWST, Euclid and ELTs. In particular, \mistralS{} will be an invaluable tool for exploring the origin of high-redshift massive galaxy quenching, a phenomenon increasingly observed with JWST \citep{Valentino2023,Carnall2023,deGraaff2025}, but not yet fully captured by current cosmological simulations \citep[e.g.][]{Weller2025}.

\section*{Acknowledgements}

We thank the anonymous referee for providing helpful comments which improved this paper. This work benefited from the "Learning the Universe" collaboration, supported by the Simons Foundation. The simulations presented in this paper have been partly performed using the facilities of the Scientific IT and Application Support Center (SCITAS) of EPFL, using the cluster Jed. This work was also supported by a grant from the Swiss National Supercomputing Centre (CSCS) under project ID s1231. The analysis of the simulations were performed using the Yggdrasil machine, from the Baobab HPC service of the University of Geneva. We are grateful to Volker Springel and the TNG collaboration team for providing access to the \arepo{} code and to the physical modules used in the TNG simulations. We additionally thank Kung-Yi Su for sharing the initial conditions of the idealised galaxy used in this paper, Tiago Costa for his comments on the original draft, and Yohan Dubois and Jenny Greene for instructive discussions. MF and MH acknowledge funding from the Swiss National Science Foundation (SNF) via a PRIMA Grant PR00P2 193577 “From cosmic dawn to high noon: the role of black holes for young galaxies”. EC was supported by the National Research Foundation of Korea (NRF-RS-2023-00213322). TN acknowledges support from the Deutsche Forschungsgemeinschaft (DFG, German Research Foundation) under Germany’s Excellence Strategy - EXC-2094 - 390783311 from the DFG Cluster of Excellence "ORIGINS". RW acknowledges funding of a Leibniz Junior Research Group (project number J131/2022).

The main roles of the authors were, using the CRediT (Contribution Roles Taxonomy) system\footnote{https://credit.niso.org}: \textbf{Marion Farcy:} conceptualization; data curation; formal analysis; investigation; methodology; visualization; writing -- original draft, \textbf{Michaela Hirschmann:} conceptualization; funding acquisition; methodology; resources; writing - review and editing, \textbf{Rachel S. Somerville:} conceptualization; methodology; writing - review and editing, \textbf{Ena Choi:} conceptualization; methodology, \textbf{Sophie Koudmani:} conceptualization; methodology; writing - review and editing, \textbf{Thorsten Naab:} conceptualization; methodology; writing - review and editing, \textbf{Rainer Weinberger:} methodology; writing - review and editing, \textbf{Jake S. Bennett:} writing - review and editing, \textbf{Aklant K. Bhowmick:} writing - review and editing, \textbf{Hyunseop Choi:} writing - review and editing, \textbf{Lars Hernquist:} writing - review and editing, \textbf{Julie Hlavacek-Larrondo:} writing - review and editing, \textbf{Bryan A. Terrazas:} resources; writing - review and editing, \textbf{Francesco Valentino:} writing - review and editing.



\section*{Data Availability}
The data underlying this article will be shared on reasonable request to the corresponding author. The \mistral{} model has primarily been developed by MF, MH, RSS, EC, SK and TN. We kindly ask that any inquiries regarding the use of \mistral{} or requests for collaboration be directed to these authors. The IllustrisTNG simulations are publicly available and accessible at \url{www.tng-project.org/data} \citep{Nelson2019}.



\bibliographystyle{mnras}
\bibliography{biblio}




\appendix
\section{Testing Mistral at higher resolution}
\label{app:res}

In order to study the resolution convergence of the results discussed in Section~\ref{section:resultsz2}, we re-simulate 3 of the 15 galaxies of our sample with an increased resolution. Specifically, we target halos with group number 61, 30 and 10. We construct their new ICs at two times higher spatial resolution, i.e. we increase the number of particles by a factor of 8, which roughly corresponds to TNG50 resolution (as summarised in Table~\ref{tab:run_res}). To limit their computational cost, we run these higher-resolution simulations down to $z=3$. We refer to them as the $\rm res\times2$ simulations, as opposed to the $\rm res\times1$ simulations studied in Section~\ref{section:resultsz2}, that adopt the resolution of their parent TNG100 simulation. Apart from resolution, both sets of simulations use the same galaxy formation physics, as described throughout the paper. We only increase the weighted number of gas cells in the BH smoothing volume $n_{\rm BH,ngb}$ by a factor of two, as was originally done from TNG100 to TNG50 (and as explained by \citealp{Weinberger2017}).

\begin{table}
	\centering
	\caption{Simulation resolution as a function of the zoom factor, with $\rm res\times1$ corresponding to TNG100 resolution and $\rm res\times2$ to (roughly) TNG50 resolution. From top to bottom: $m_{\rm DM}$: dark matter mass resolution; $m_{\rm baryons}$: mean baryon mass resolution; $\rm \epsilon_{DM,stars}$: Plummer equivalent gravitational softening of the collisionless component; $\rm \epsilon_{gas}$: minimum value of the gravitational softening of the gas; $n_{\rm BH,ngb}$: weighted number of gas cells in the BH smoothing volume; $z_{\rm end}$: redshift reached at the end of the run, for the last snapshot.}
	\label{tab:run_res}
	\begin{tabular}{lccc}
        Resolution & \vline & TNG100 ($\rm res\times1$) & $\sim$ TNG50 ($\rm res\times2)$\\
        \hline
        $m_{\rm DM}\,\rm[M_{\odot}]$ & \vline & $7.5\times10^6$ & $9.4\times10^5$\\
        $m_{\rm baryons}\,\rm[M_{\odot}]$ & \vline & $1.4\times10^6$ & $1.7\times10^5$\\
        $\rm \epsilon_{DM,stars}\,\rm[cpc/h]$ & \vline & $1000$ & $500$\\
        $\rm \epsilon_{gas}\,\rm[cpc/h]$ & \vline & $125$ & $62.5$\\
        $n_{\rm BH,ngb}$ & \vline & $256$ & $512$\\
        $z_{\rm end}$ & \vline & $2$ & $3$\\
        \hline
	\end{tabular}
\end{table}

\begin{figure}
    \centering
    \includegraphics[width=\columnwidth]{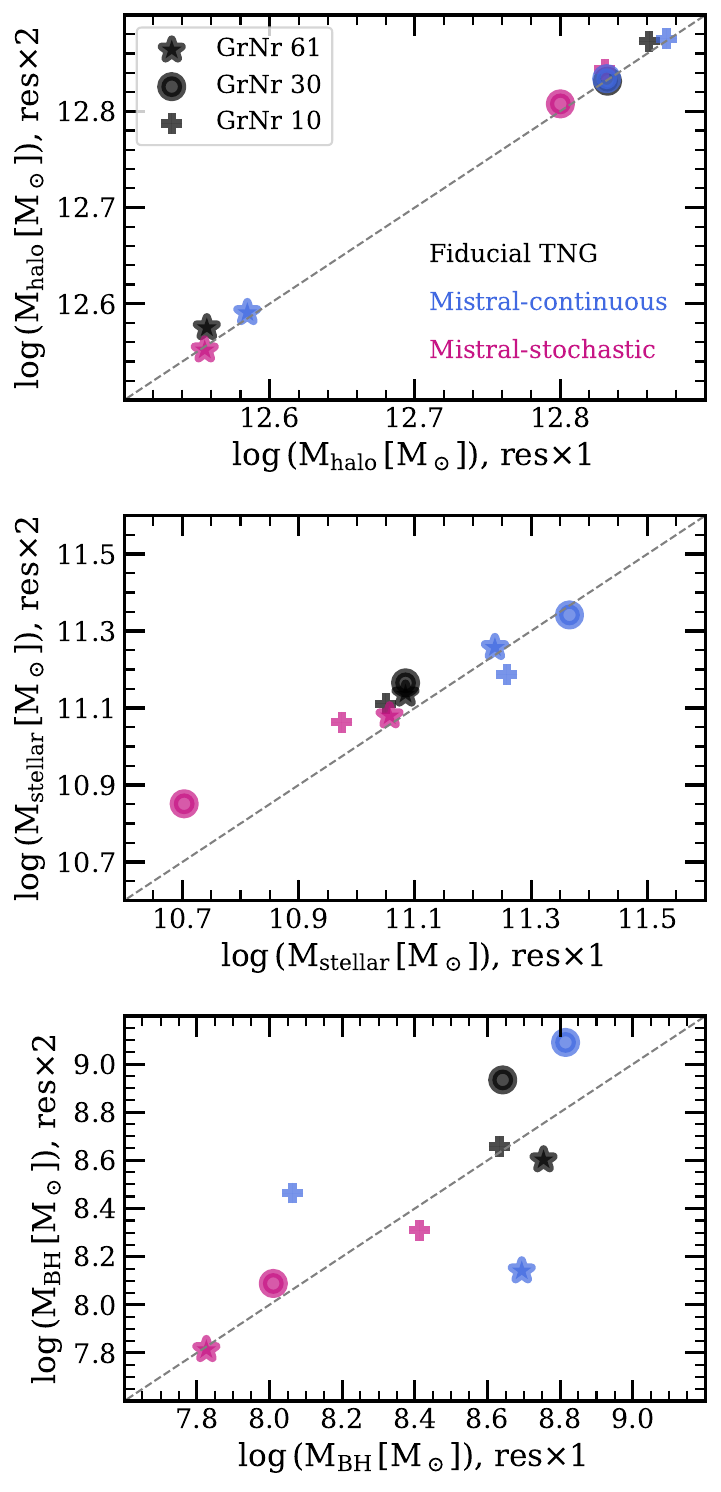}
    \caption{Halo mass, stellar mass and black hole mass at $z=3$ for three galaxy simulations at TNG100 (res$\times1$) and TNG50 (res$\times2$) resolution. Zoom simulations with the \tng{} physics, \mistralC{} and \mistralS{} are shown in black, blue and purple, respectively. In each panel, the dashed line shows the 1:1 relation.}
    \label{fig:mres}
\end{figure}

Fig.~\ref{fig:mres} shows the halo, stellar and black hole masses in the $\rm res\times2$ versus the $\rm res\times1$ simulations. We compare these three properties at $z=3$ for zoom simulations using \mistralC{} (blue), \mistralS{} (purple), as well as the \tng{} physics (black), for reference. Halo and stellar masses are fairly well converged with these three setups, as they follow closely the one-to-one relation shown in dashed lines. While \mistralS{} leads to similar BH masses regardless of resolution, \mistralC{} and the \tng{} model can both lead to more and less massive BHs when changing resolution, and this effect is more pronounced with \mistralC{}. Resolution convergence in cosmological simulations is a long standing issue, due to the complex and non-linear nature of the interactions between feedback processes and galaxy evolution. In order to recover similar evolution of stellar and black hole masses with time, it may be necessary to recalibrate the efficiency of AGN feedback against $z=0$ observations when changing resolution, which we will address in a future work.

\section{Calibration of the wind efficiency parameter for Mistral}
\label{app:calib}

Our \mistral{} model contains two free parameters: the average wind velocity $v_{\rm w}$, and the AGN feedback efficiency $\epsilon_{\rm w}$. Throughout this paper, we keep $v_{\rm w}=10^4\,\rm km\,s^{-1}$, as done by \citet{Choi2012,Choi2014,Choi2015,Choi2017,Choi2018}. In order to determine which $\epsilon_{\rm w}$ should be used in our cosmological zoom simulations, we explore the impact of varying this parameter on 3 of the 15 galaxies from our sample. More specifically, we focus on galaxies hosted in the halos with group number 631, 123 and 57 at $z=2$ (snapshot 33 from TNG100), which correspond to GrNr = 692, 101 and 64 at $z=0$ (snapshot 99 from TNG100, from which their ICs are extracted). We simulate these three halos with \mistralC{} and \mistralS{}, down to $z=0$, and with $\epsilon_{\rm w}=5\times10^{-4}$, $10^{-3}$ and $5\times10^{-3}$. 

\begin{figure*}
    \centering
    \includegraphics[width=\textwidth]{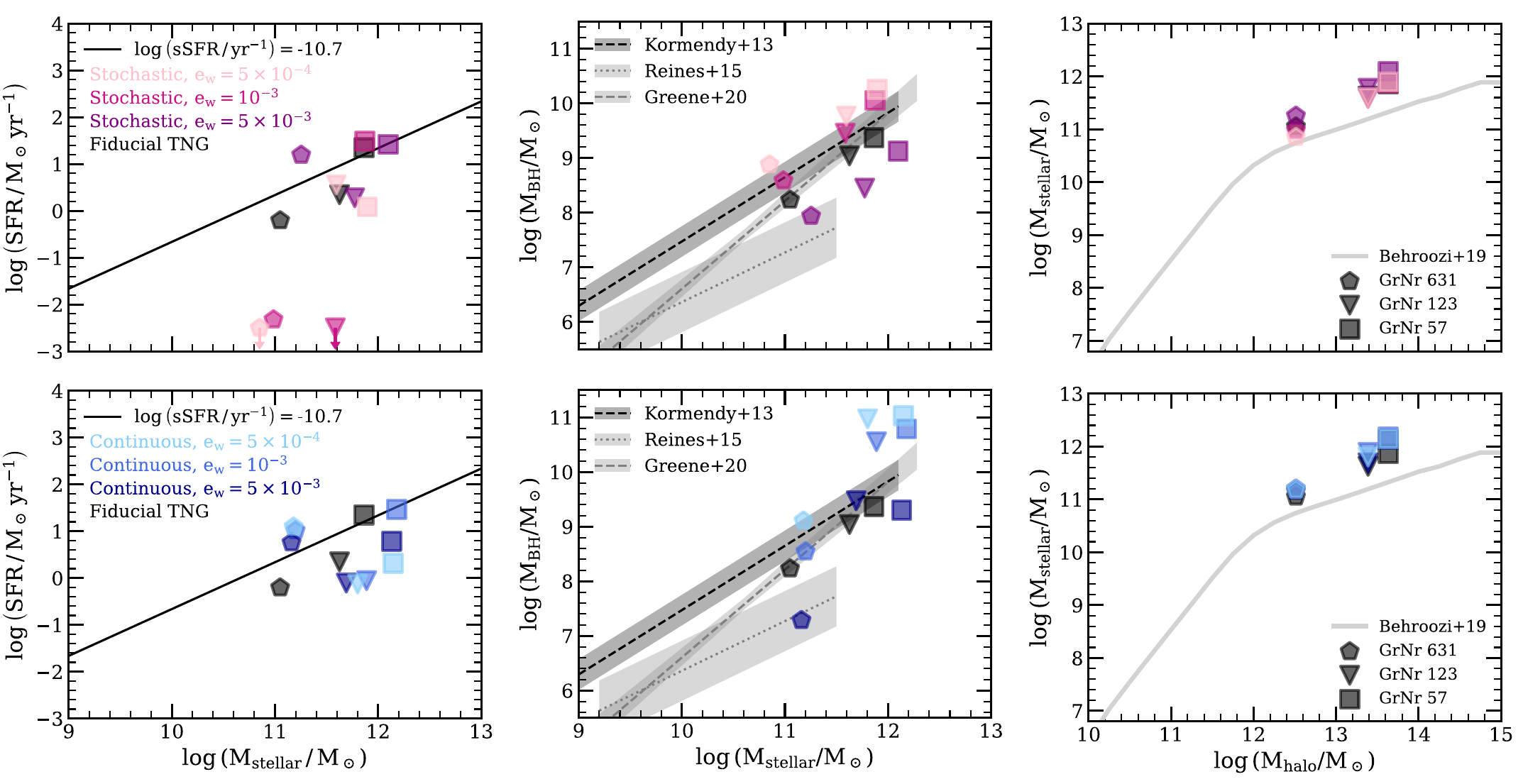}
    \caption{Scaling relations at $z=0$ for zoom simulations with \mistralS{} in the upper row (respectively \mistralC{} in the bottom row), with $\epsilon_{\rm w} = 5\times10^{-4}, 10^{-3}, 5\times10^{-3}$ in pink (light blue), purple (blue) and dark purple (dark blue). For reference, we show the counterpart zoom simulations with the \tng{} physics in black. From left to right: star formation rate and BH mass versus stellar mass (with observational relations from \citealp{Kormendy&Oh2013,Reines&Volonteri2015,Greene2020}), and stellar mass to halo mass relation (displayed at the same x-coordinate, compared with the empirical relation from \citealp{Behroozi2019}). The arrows in the upper left plot denote a SFR lower than the limit of the plot. The legend with the markers indicates the halo group number at $z=2$, for consistency with the rest of the paper.}
    \label{fig:mscalingz0}
\end{figure*}

\begin{figure*}
    \centering
    \includegraphics[width=\textwidth]{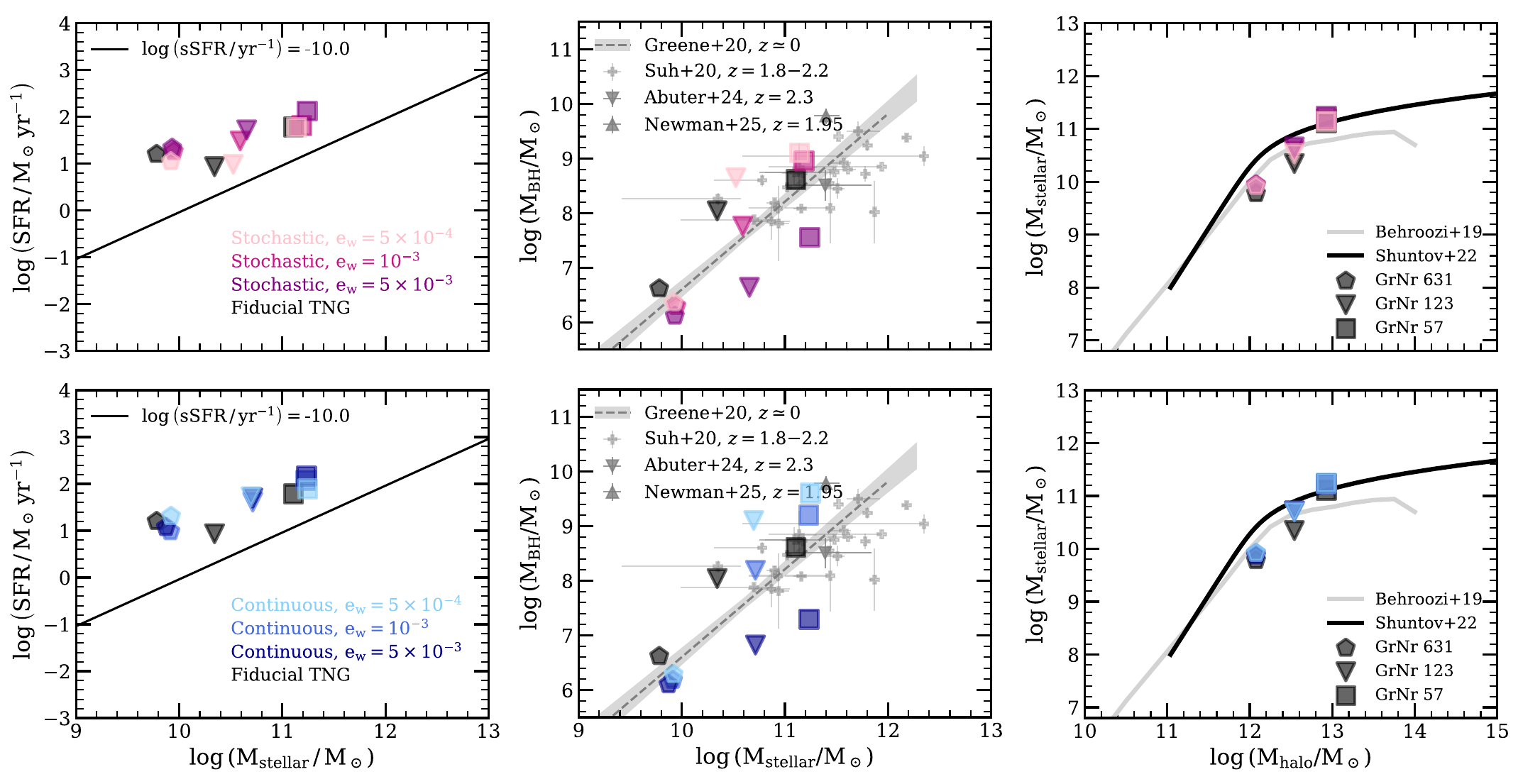}
    \caption{Same as Fig.~\ref{fig:mscalingz0} but at $z=2$, with observational estimates from \citet{Suh2020,Abuter2024,Newman2025} in the middle panels and from \citet{Shuntov2022} in the right ones.}
    \label{fig:mscalingz2}
\end{figure*}

In Fig.~\ref{fig:mscalingz0} and \ref{fig:mscalingz2}, we show the results when increasing the value of $\epsilon_{\rm w}$ from light to darker shades of blue and purple, for \mistralC{} and \mistralS{} respectively. For reference, we also re-run these three objects using the zoom-in technique with the \tng{} physics, which is displayed with black symbols. From left to right, the two figures show SFR and BH mass versus stellar mass, and the stellar mass to halo mass relation at $z=0$ (Fig.~\ref{fig:mscalingz0}) and $z=2$ (Fig.~\ref{fig:mscalingz2}). Overall, and with both versions of \mistral{}, increasing $\epsilon_{\rm w}$ leads to smaller black hole masses and to slightly higher stellar masses. This is because the fraction of gas mass accreted by the BH decreases when $\epsilon_{\rm w}$ increases (while the fraction of gas mass receiving BH energy increases, see also Fig.~\ref{fig:ew-vw}), which reduces the total amount of BH energy released and, as a result, reduces the ability of AGN feedback to impact star formation. For this reason, galaxies tend to have lower SFRs and to be more easily quenched at $z=0$ in the runs with $\epsilon_{\rm w}=5\times10^{-4}$. In this paper, we adopt $\epsilon_{\rm w}=10^{-3}$ as the best compromise to produce realistic black hole and stellar masses at $z=0$ and $z=2$, simultaneously suppressing star formation at $z=0$ for most simulations. A higher (respectively lower) value of $\epsilon_{\rm w}$ could have been used with \mistralS{} (\mistralC{}), but we prefer to use the same value in both models. In particular, this choice is motivated by the fact that $\epsilon_{\rm w}=5\times10^{-3}$ with \mistral{} produces low BH masses at $z=2$, which would make it less relevant for studying the impact of AGN-driven winds at cosmic noon.

\bsp	
\label{lastpage}
\end{document}